\documentclass[referee]{aa}
\usepackage[varg]{txfonts}
\usepackage{graphicx}
\usepackage{natbib}
\usepackage{pifont}
\usepackage{xcolor}
\bibpunct{(}{)}{;}{a}{}{,} 

\begin{document}

\title{Characterisation of chaos in meteoroid streams}
\subtitle{Application to the Taurids}
\author{Ariane Courtot \inst{1,2} \corrauth{ariane.courtot@obspm.fr} \and Melaine Saillenfest \inst{3,4,2} \and Marc Fouchard \inst{2} \and Jérémie Vaubaillon \inst{2}}
\institute{
Astronomical Institute, Ond\v rejov Observatory, Fri\v cova 298, 25165 Ond\v rejov, Czech Republic
\and
LTE, Observatoire de Paris, PSL Research University, Sorbonne Universit\'{e}, Université de Lille, LNE, CNRS, France
\and
Department of Earth Sciences, National Taiwan Normal University, Taipei 116, Taiwan
\and
Center of Astronomy and Gravitation, National Taiwan Normal University, Taipei 116, Taiwan
}
\date{Received September 2026 / Accepted date }

\abstract
{} 
{We explore the dynamics of the Northern and Southern Taurid meteoroid streams to identify major mechanisms structuring the evolution of the streams.} 
{We use chaos maps and numerical simulations to explore the dynamics of the streams. We also perform some semi-analytical analysis to confirm the mechanisms we identified in the maps.} 
{We show how Northern Taurids are structured by the von Zeipel-Lidov-Kozai mechanism and how it interacts with several mean-motion resonances with Jupiter. This mechanism is visible as an increase in inclination for most particles, which is less prominent for those trapped in mean-motion resonances. It is especially strong for larger initial semi-major axis within the stream. We specifically study the 7:2 mean-motion resonance with Jupiter to confirm it leads to observable swarms. Furthermore, we explore the influence of non-gravitational forces, showing that they blur the influence of mean-motion resonances but not the von Zeipel-Lidov-Kozai mechanism. We also show that the dynamics of the Southern Taurids are very similar to the Northern Taurids.} 
{} 


\maketitle

\section{Introduction}

Meteoroid steams are created when small particles are ejected from a comet or an asteroid. Due to their very nature, the dynamics of these streams are complex: close encounters with planets and effect of non-gravitational forces, for example, combine with the effect of resonances, and secular dynamical mechanisms. Understanding the dynamics of meteoroid streams is instrumental in answering one of the main questions in meteor science today: to which level can we reliably identify the parent body of the meteoroids we detect on Earth as meteors? Depending on the answer to this question, the study of meteors on Earth would help us to understand better asteroids and comets, as meteors would be markers and indicators of small bodies origin, dynamics and/or composition. 

Meteor showers are defined as a set of meteors with similar orbits coming from one single body, while we call "meteor group" a set of meteors with similar orbits that might or might not come from one single parent body. Meteor showers are meteor groups whose common origin was proved definitely. Understanding the dynamics of meteoroid streams could also help differentiate which meteors belong to a serendipitous meteor group and which belong to a true meteor shower.

In previous studies \citep{Courtot_al_2023, Courtot_al_2024}, we have successfully used chaos maps to identify dynamical mechanisms playing a role for specific meteoroid streams (the Geminids, the Draconids and the Leonids). This study of the Taurids allows us to work on two problems that we did not study in these previous works:~parenthood of the streams and identification of a general dynamical mechanism.

The Taurids is a complex set of meteors. Usually, papers cite the comet 2P/Encke as the parent body of the Taurids, but this has been disputed. In fact, it seems the Taurids parent body could be a progenitor of 2P/Encke, rather than the comet itself \citep{Devillepoix_al_2021,Egal_al_2021,Egal_al_2022}. This discussion on the parent body of the Taurids means their "meteor shower" status could be reevaluated, contrarily to the Geminids, Draconids and Leonids. Thus, it is a good candidate to start probing the difference between meteor showers and meteor groups on a dynamical level. In other words, the difficulty in identifying the Taurids as a meteor shower might be explained by an underlying dynamical mechanism that we aim here to search for.

The three meteoroid streams we have already studied have very different orbits from one another, but the main dynamical mechanism we identified is the same: one (or more) mean-motion resonance with a planet $P$ captures meteoroids and prevents them from encountering $P$, keeping them on regular orbits, whereas other particles out of the resonance do experience close encounters, which raise the chaoticity of their orbit. The Taurids orbit is different from all of the streams we have studied before, both in semi-major axis and inclination, which makes it a good candidate to test if this mechanism is general to all streams we observe on Earth.

The Taurids is sometimes called "the Taurids complex": it is composed of several sub-streams, which might all come from the same parent body, or not. In order to restrict our study to a manageable amount of sub-streams, we chose to focus on the Northern Taurids (NTA) and Southern Taurids (STA). They are well-known and recognised as the two main sub-streams of the Taurid complex. They are also listed as two separate established meteor showers in the International Astronomical Union Meteor Database (IAU MDC): they are usually considered as both coming from the same parent body, but with different dynamical evolutions that separated them in two different streams.

In the next section, we present our method, which is similar to the one we used to analyse the Geminids, Draconids and Leonids. Then, in Sect.~\ref{sec:res}, we show the results obtained, focusing on relatively big particles from the NTA, before widening the results to smaller NTA particles and to STA particles. We explore how close encounters, mean-motion resonances and the von Zeipel–Lidov–Kozai mechanism interact with the particles. We also analyse the effect of non-gravitational forces. Finally, we conclude.

\section{Method}\label{sec:meth}

For each integration, 99720~particles are integrated for 1000~years. The integrator used is the Radau, a 15th-order integrator with an adaptative time step. Gravitational forces are computed using the positions of the planets given by the ephemerides INPOP \citep{Fienga_al_2009}, while several non-gravitational forces are taken into account (solar radiation pressure and Poynting-Robertson drag). The particles have all the same bulk density (1000~kg~m$^{-3}$). We define as close encounter the situation where a particle is closer to a planet than its Hill radius. In case of such a close encounter, we tag this particle as having had a close encounter with a given planet. For a more extensive description of the method, see \citet{Courtot_al_2023}

The chaos indicator chosen is the Orthogonal Fast Lyapunov Indicator (OFLI), following the study of chaos indicator performed in \citet{Courtot_al_2023}. Its definition can be found in \citet{Fouchard_al_2002}.

To create the initial conditions, we randomly select orbital elements centred on the IAU MDC definition of the NTA and STA, following a uniform distribution. Table~\ref{tab:IC} shows which limits we chose for each orbital elements. The initial time is the year 1100 CE.

\begin{table}
    \caption{Ranges of initial orbital elements}
    \label{tab:IC}
    \centering
    \begin{tabular}{c c c}
    \hline \hline
          & NTA & STA \\
         \hline
         $a$ (au) & 1.542 - 2.742 & 1.569 - 2.569 \\
         $e$ & 0.765 - 0.865 & 0.711 - 0.861 \\
         $I$ (deg) & 1.7 - 3.7 & 4.7 - 6.7 \\
         $\Omega$ (deg) & 208.3 - 264.3 & 20.35 - 83.35 \\
         $\omega$ (deg) & 276.1 - 303.1 & 85.6 - 123.6 \\
         \hline
    \end{tabular}
    \tablefoot{The range of mean anomaly for both streams is 0 - 360\degr.}
\end{table}

We integrated particles with radius between 10 and 100~mm from the NTA stream (hereafter "NTA BIN10100"), particles with the same size from the STA stream (hereafter "STA BIN10100") and finally particles with smaller sizes, and especially particles with radius between 0.01 and 0.1~mm from the NTA stream (hereafter "NTA BIN00101"). The non-gravitational forces only play a role for particles smaller than 10~mm, allowing us to analyse separately the effect of these forces. We cannot select a radius smaller than 0.01~mm because, for such small particles, the non-gravitational forces are heavily modified. In fact, at smaller bins of radii, the radiation-matter interaction passes in another regime. Furthermore, meteoroids producing visible meteors are usually of sizes between 0.01~and 100~mm.

During the integration, particles that reach a distance to the Sun lower than 0.02~au or higher than 1000~au are taken out of the computation. The first case corresponds to a particle that will crash into the Sun in a few orbits, while the second is characteristic of a particle ejected out of the Solar System.

\section{Results}\label{sec:res}

\begin{figure}
    \includegraphics[scale = 0.5]{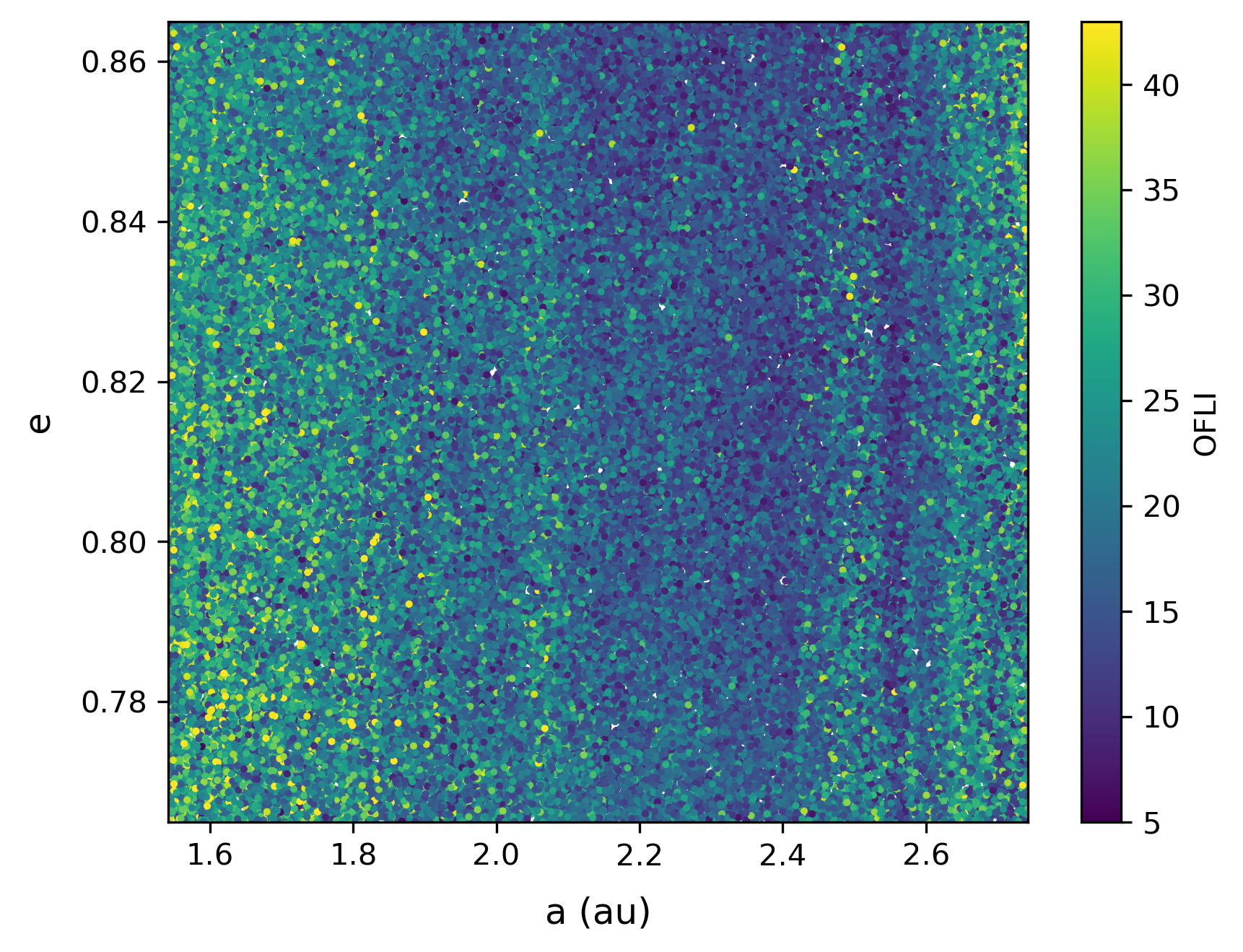}
    \includegraphics[scale = 0.5]{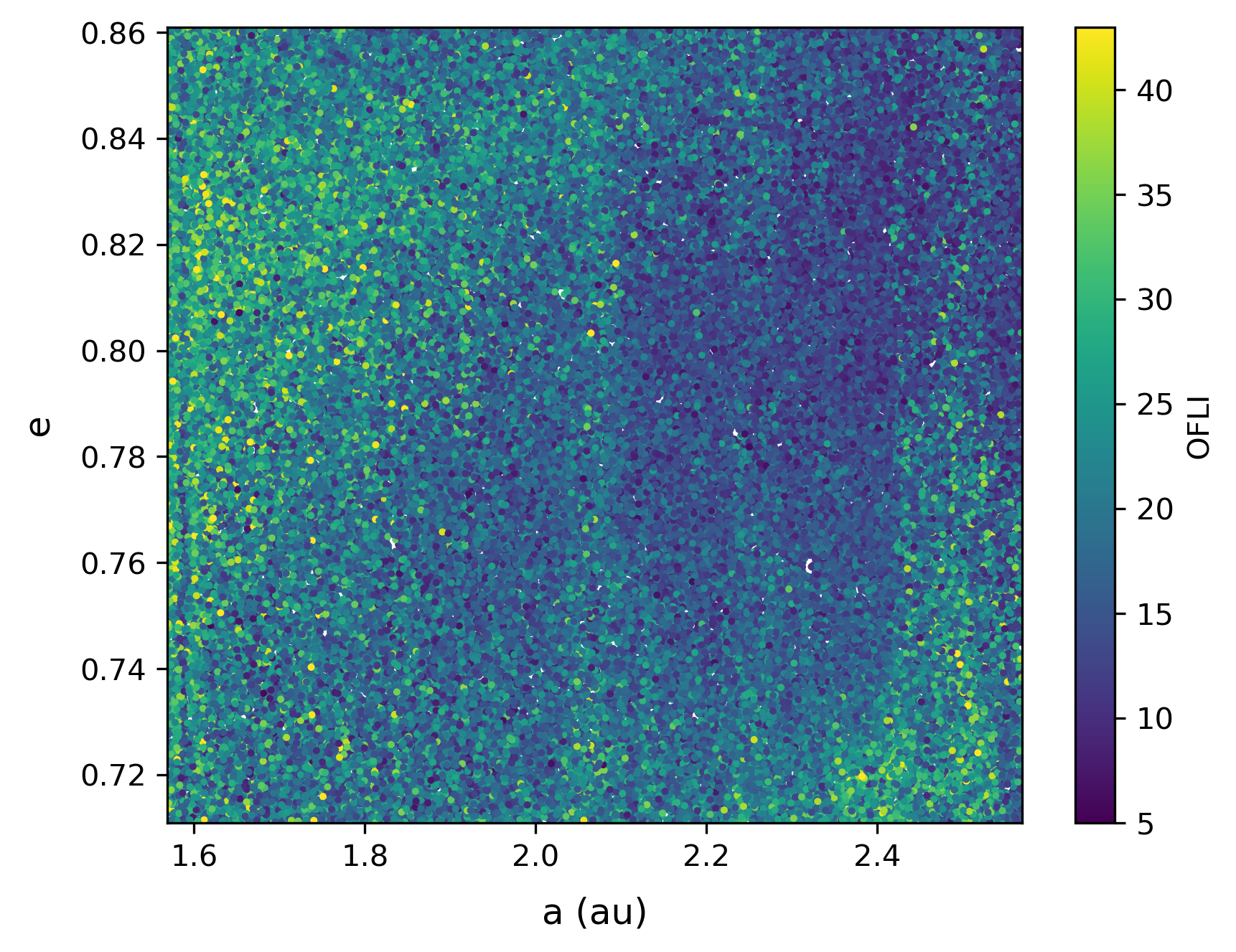}
    \caption{Semi-major axis and eccentricity maps drawn from the data sets NTA BIN10100 (top) and STA BIN10100 (bottom). $a$ is the initial semi-major axis, $e$ the initial eccentricity and the colorbar represents OFLI after 1000~years integration.}
    \label{fig:NTA_STA}
\end{figure}

While we created maps using all initial orbital elements combined with the final OFLI, we will focus on maps using the initial semi-major axis, initial eccentricity and final OFLI as these are the best representatives of our results. Fig.~\ref{fig:NTA_STA} shows the chaos maps obtained for the NTA BIN10100 and STA BIN10100 initial conditions.

We chose to restrict the colour scale of the OFLI, plotting all particles above 43 at the same colour. For the NTA BIN10100, only 0.32\% particles have a OFLI above this value, with a maximum of 63.05. For the STA BIN10100, 0.26\% particles OFLI reach above 43, with a maximum of 68.09. This restriction allows for a clearer map.

Comparing both maps of Fig.~\ref{fig:NTA_STA}, we see that they are qualitatively similar, which means that most analyses of the NTA will also apply to the STA. Thus we will present in this section our analysis of the NTA first before verifying if our conclusions can be applied to the STA as well.

We note here that no particle was taken out of the integration regarding the NTA BIN10100 dataset. In the case of the STA BIN10100, only one particle was ejected out of the Solar System, and is therefore not included in the maps.

\subsection{Sources of chaos}\label{subsec:mmr}

The maps show areas of chaos depending on the initial semi-major axis (see the vertical structures in Fig.~\ref{fig:NTA_STA}). Thus we looked for mechanisms that could bring chaos to an otherwise quite regular dynamic. We first analysed the effect of close encounters, a known source of chaos, and we also started to analyse mean-motion resonances, which caused more regular dynamics in previous works. Finally, we had to explore other mechanisms.

\subsubsection{Close encounters}

\begin{figure}
    \includegraphics[scale = 0.5]{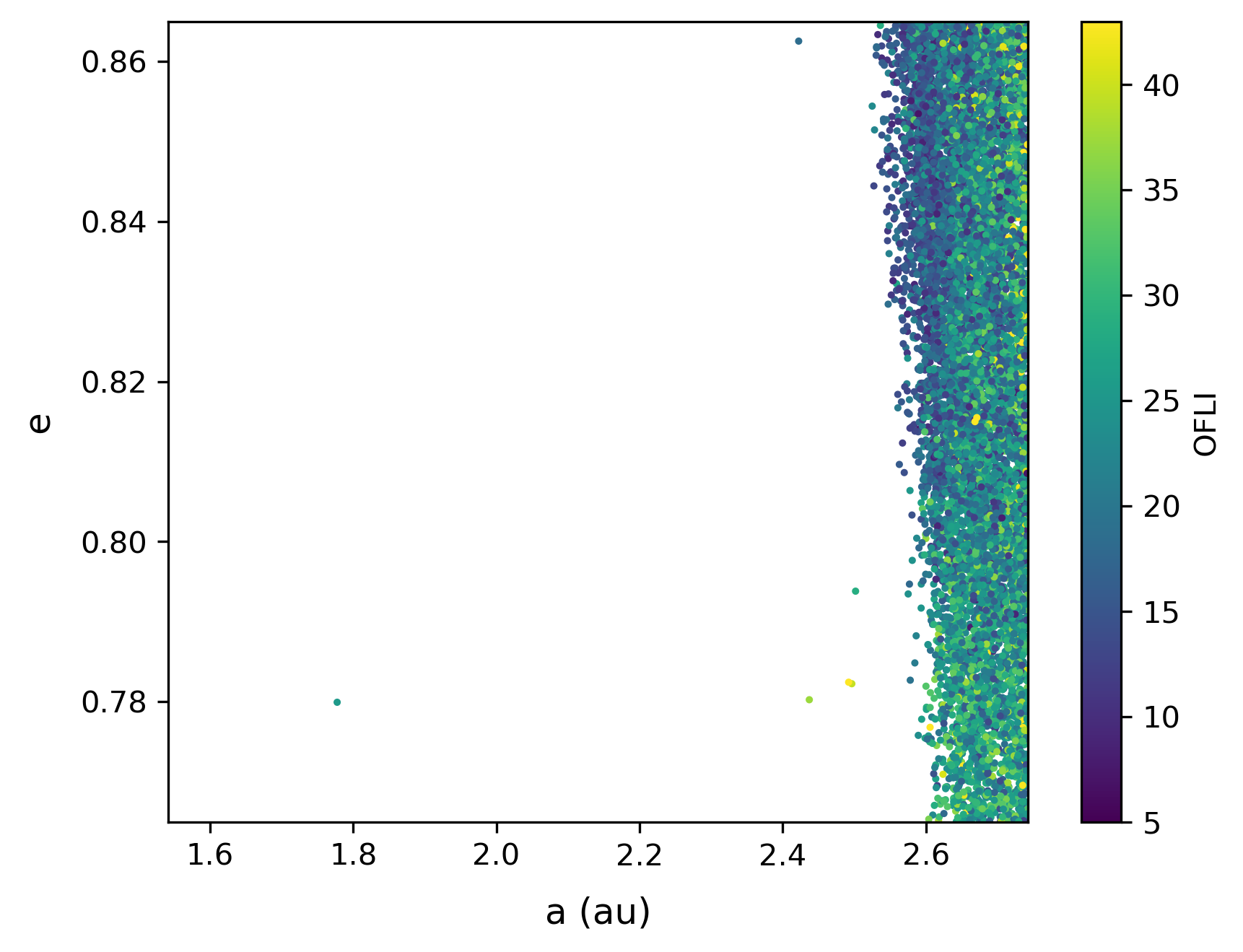}
    \caption{Semi-major axis and eccentricty chaos maps for the NTA BIN10100, but only the particles encountering Jupiter are plotted. The entire area of the initial conditions is accounted for, for better comparison with the other maps presented in this paper.}
    \label{fig:NTA_rencJup}
\end{figure}

Fig.~\ref{fig:NTA_rencJup} shows the same semi-major axis and eccentricity chaos map, but only the particles that encounter Jupiter are plotted. The close encounters with Jupiter explain very well the chaos of the orbits with an initial semi-major axis higher than 2.6~au. 

\begin{figure}
    \includegraphics[scale = 0.5]{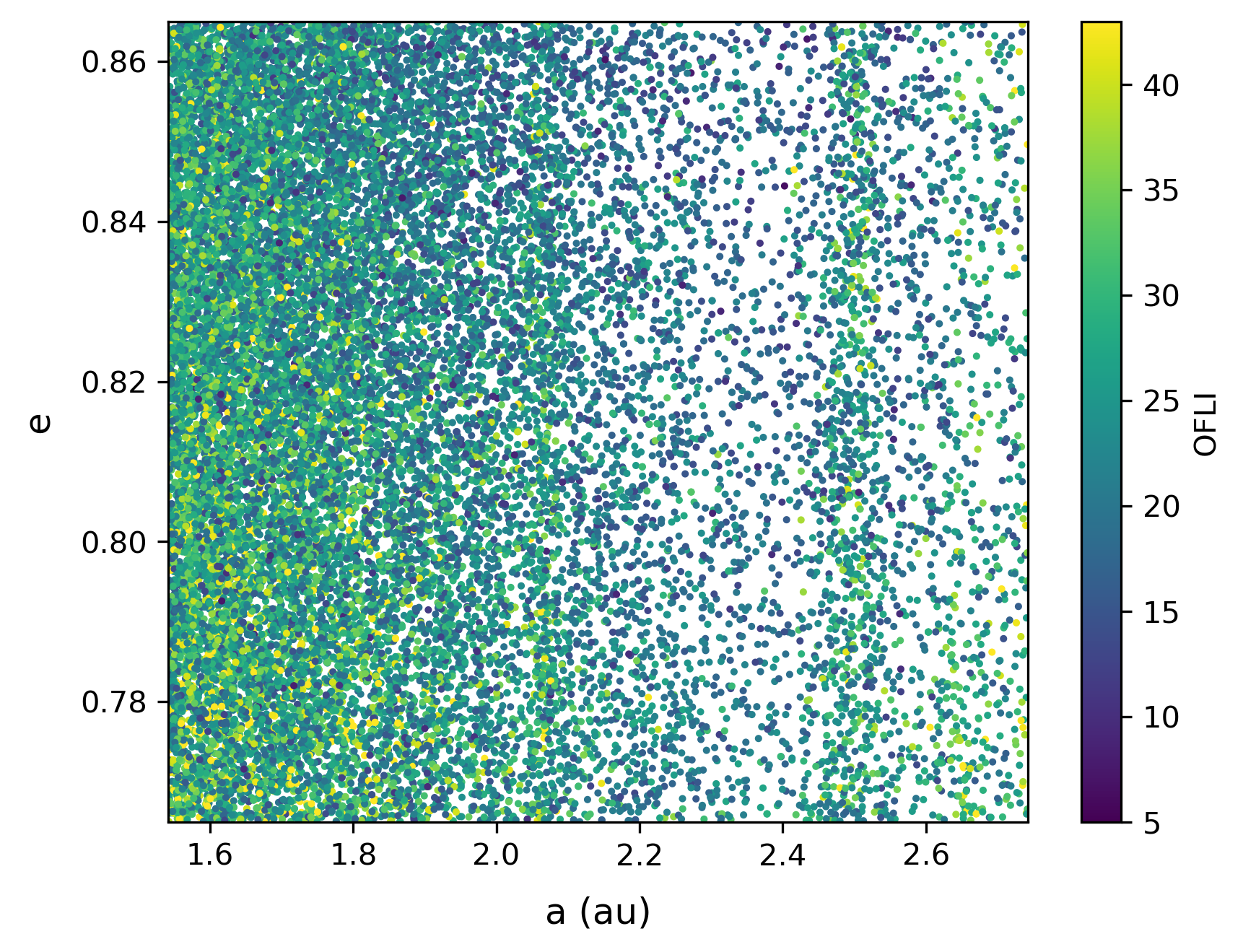}
    \caption{Similar to Fig.~\ref{fig:NTA_rencJup} but the particles represented encountered Venus and/or Mars and/or the Earth.}
    \label{fig:NTA_inners}
\end{figure}

We also analysed the close encounters with the inner planets. Fig.~\ref{fig:NTA_inners} shows particles that encountered Venus, Mars or Earth (there is not enough encounters with Mercury to make a difference). A majority of these encounters are with Venus and the Earth; they also bring more chaos than encounters with Mars. Most of the chaos brought by encounters from these three planets are to the orbits with initial semi-major axis lower than 1.8~au. There are also two lines with much more encounters at about 2.06~au and 2.5~au. We will explore what creates such lines in further sections.

\subsubsection{Mean-motion resonances}

\begin{figure}
    \includegraphics[scale = 0.5]{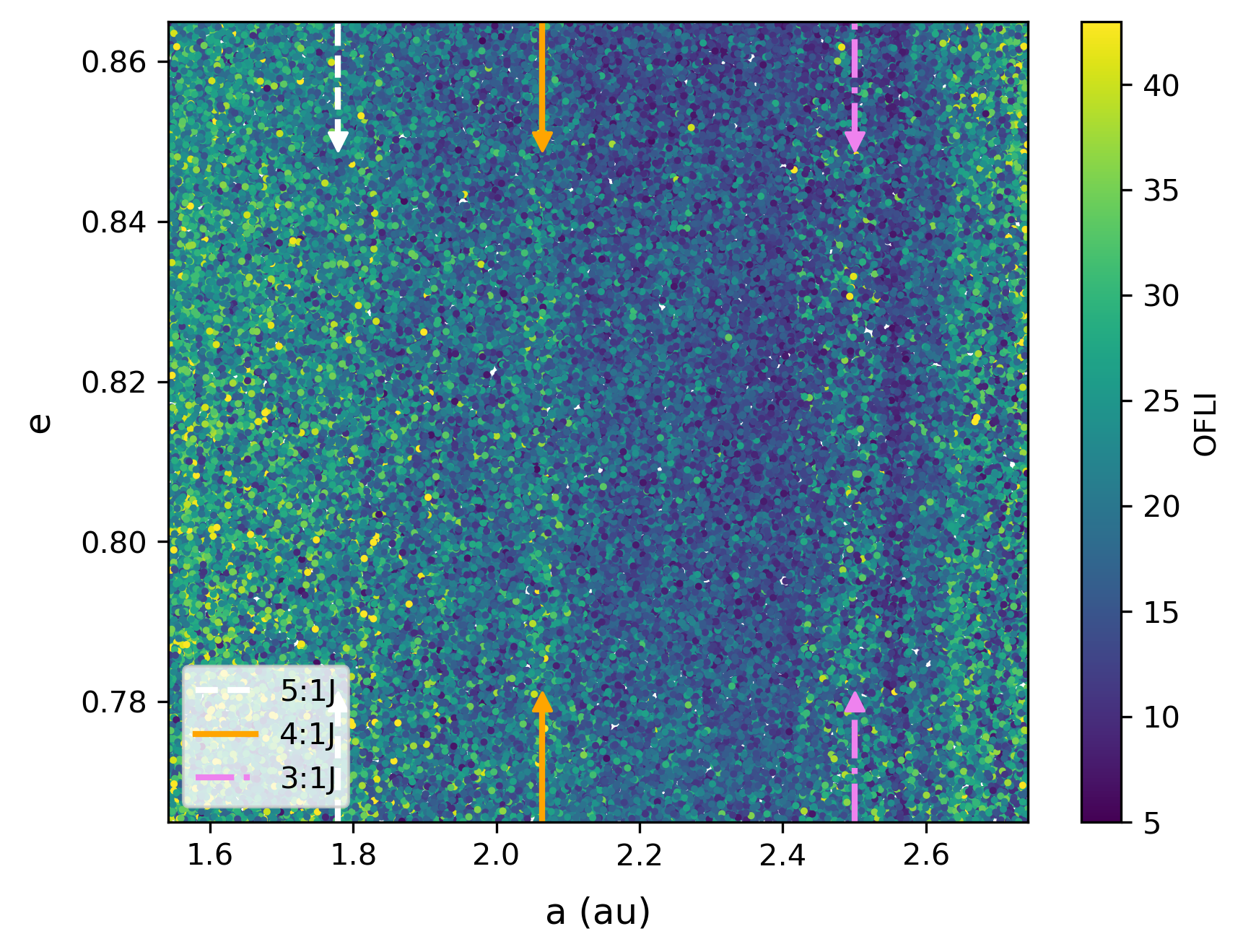}
    \caption{Semi-major axis and eccentricity map of the NTA BIN10100. The semi-major axis location of three MMR with Jupiter are indicated by arrows.}
    \label{fig:NTA_reso}
\end{figure}

In this last map (Fig.~\ref{fig:NTA_inners}), the two close encounters lines detected fit very well with chaotic lines. These lines also fit perfectly with some mean-motion resonances (MMR). These MMR are always with Jupiter. The precise theoretical MMR position are listed in Table~\ref{tab:MMR}. Fig.~\ref{fig:NTA_reso} shows the MMR identified superimposed on a NTA map. 

It should be noted here that we are confirming the importance of two MMR in the dynamics of Taurid particles already identified in a previous study. Indeed, \citet{Valsecchi_al_1995} showed that the 4:1 MMR with Jupiter, as well as the 3:1 with Jupiter, play a role in the dynamics of the Taurids steam.

\begin{table}
    \caption{MMR identified in the NTA BIN101000 chaos map}
    \label{tab:MMR}
    \centering
    \begin{tabular}{c c}
    \hline \hline
         MMR & $a$ (au) \\
         \hline
         5:1 & 1.78 \\
         4:1 & 2.06 \\
         3:1 & 2.50 \\
         \hline
    \end{tabular}
\end{table}

We see in the maps that MMR appear here as sources of chaos, instead of stability, which was a surprise for us, as our previous works on meteoroid streams always revealed MMR that stabilized the particles. This means that the mechanism identified for the Geminids, Draconids and Leonids do not play a role here. The MMR affect differently the Taurids than they do the other meteoroid streams we studied, and it also means that the relatively low-level of chaoticity in the maps is linked to another mechanism.

\subsection{von Zeipel–Lidov–Kozai mechanism}\label{subsec:kozai}

We explored which mechanism(s) could maintain the particles in a regular dynamics. We started by exploring the effect of Jupiter by integrating the particles without the direct effect of Jupiter. This was achieved by omitting the computation of the effect of Jupiter gravitation on the particles. However, we did keep the ephemerides of the planets as provided by INPOP, which itself takes into account the effect of Jupiter to compute the evolution of the other planets. This means that only the direct effect of Jupiter is removed from this integration, while its indirect effect, via the other planets, is still present.

\begin{figure}
    \includegraphics[scale = 0.5]{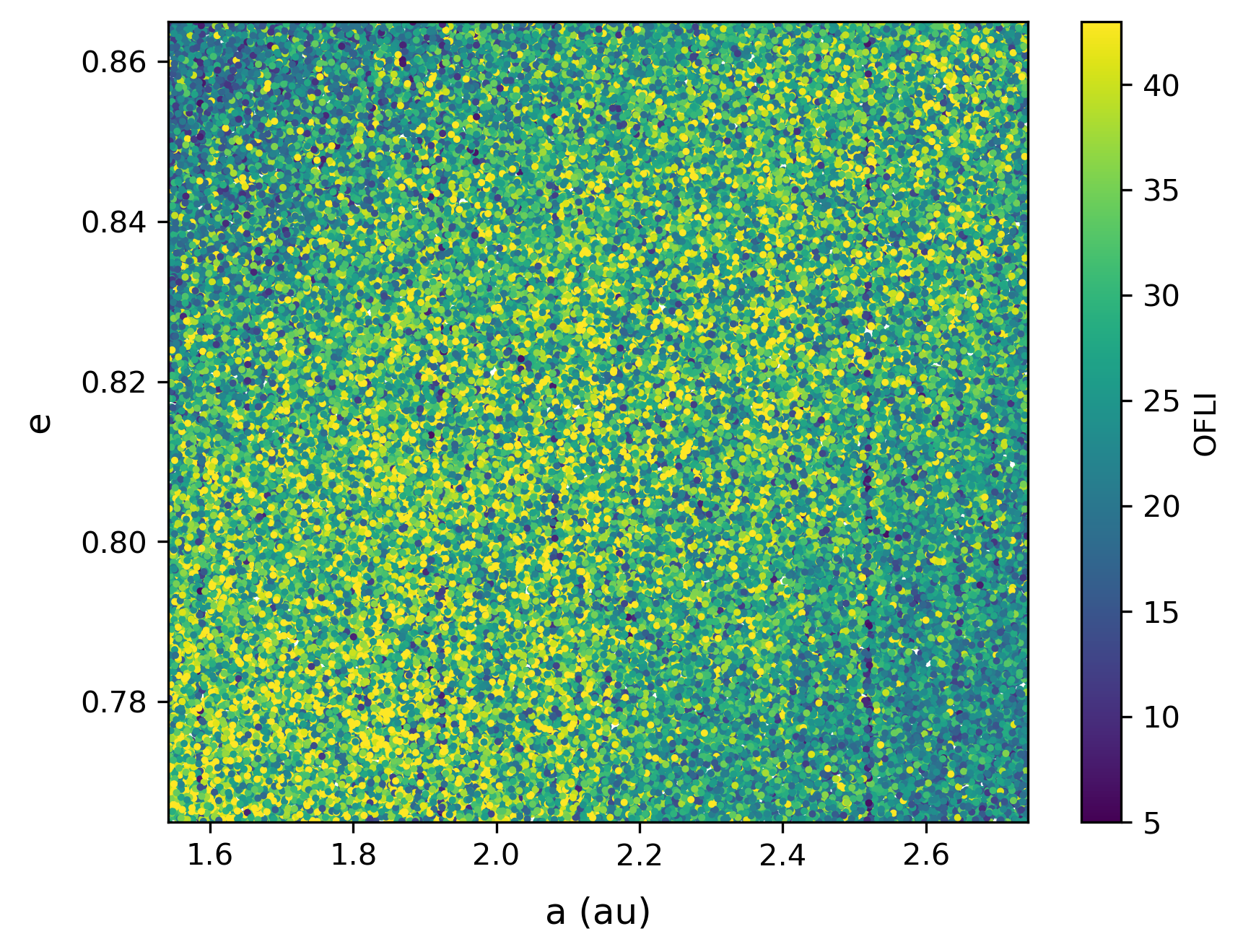}
    \caption{Semi-major axis and eccentricity map of the NTA BIN10100. This integration was performed without the direct effect of Jupiter (see text).}
    \label{fig:NTA_withoutJup}
\end{figure}

We obtained the map presented in Fig.~\ref{fig:NTA_withoutJup}. The level of chaos has risen significantly: the OFLI reaches 81 and 8.9\% of particles reach an OFLI above 43. This shows that a mechanism involving the direct gravitational effect of Jupiter is responsible for the stability. Just to corroborate these results and make sure they were not due to computational inconsistencies, we repeated the integration, this time with Jupiter and all other planets with the exception of Saturn. The map obtained is almost identical to maps with all planets included, proving the direct effect of Jupiter on the dynamics.

\begin{figure}
    \includegraphics[scale = 0.5]{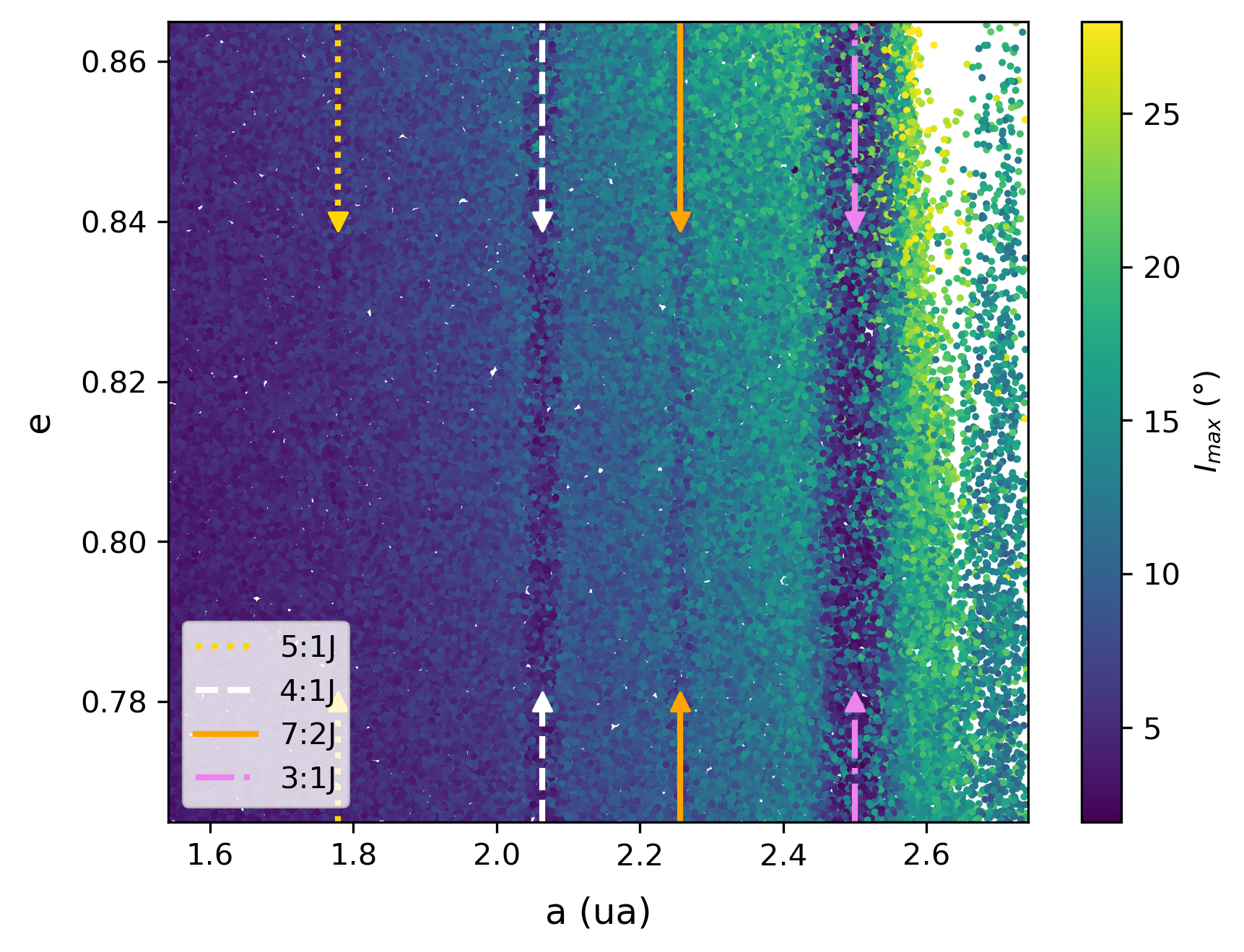}
    \caption{Initial semi-major axis, initial eccentricity and maximum inclination map of the NTA BIN10100. The only particles included are particles that did not encounter Jupiter during the integration. The values reached by the maximum inclination are between 1.92\degr and 33.27\degr.}
    \label{fig:NTA_aei}
\end{figure}

Next we explored the inclination. Fig.~\ref{fig:NTA_aei} shows the maximum inclination reached during the integration (colorbar) as a function of the initial semi-major axis and initial eccentricity. Particles encountering Jupiter during the integration are not included since their chaoticity has already been explained. When compared to Fig.~\ref{fig:NTA_STA} or Fig.~\ref{fig:NTA_reso}, it seems quite clear that a higher level of chaos is linked to a smaller maximum inclination. This makes sense: particles with orbits lying close to the ecliptic plane undergo more encounters, which lead to more chaos.

More importantly, the three MMRs already identified plus another one (7:2 with Jupiter) can also be identified. These MMRs are responsible for a higher level of chaos in the dynamics. They shield the particles from another mechanism, which tends to bring particles to a higher inclination and maintain regular dynamics.

\begin{figure}
    \includegraphics[scale = 0.5]{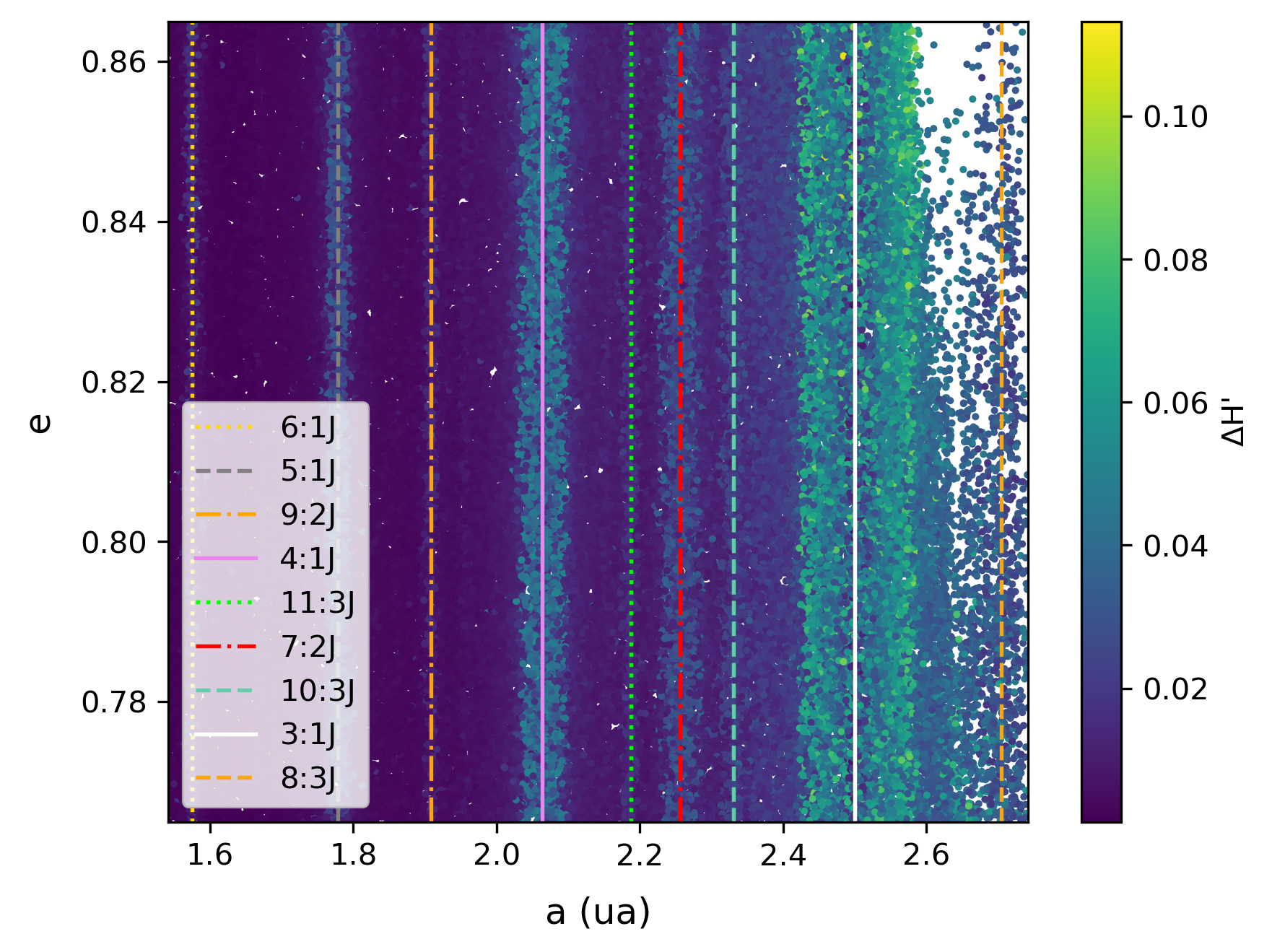}
    \caption{Difference between the maximum and minimum of $H'$ for each particle of NTA BIN10100 that do not encounter Jupiter, as a function of the initial semi-major axis and the initial eccentricity.}
    \label{fig:NTA_aeH}
\end{figure}

A natural explanation for the raise of inclination we can observe for the most regular particles would be the von Zeipel-Lidov-Kozai mechanism (vZLK; see e.g. \citealp{Lidov_1962,Kozai_1962,Kozai_1985,Thomas_Morbidelli_1996,Gomes-etal_2005,Gallardo_2006,Gallardo-etal_2012,Saillenfest-etal_2016,Ito-Ohtsuka_2019}). This secular mechanism tend to make inclination raise for particles, linking the eccentricity, inclination and argument of perihelion together. It is also linked to the direct effect of Jupiter, which would explain previous results on the direct effect of Jupiter. To verify if this might be the case here, we check whether $H$ is constant through the integration, with $H = \sqrt{a(1-e^2)} \cos I$ \citep{Thomas_Morbidelli_1996}. As the secular semi-major axis is constant during vZLK cycles outside of MMR, we actually compute: $H' = \sqrt{1-e^2} \cos I$. Fig.~\ref{fig:NTA_aeH} shows, for each particles, the difference between the maximum and minimum value of $H'$. 

The figure shows that the H' is almost constant everywhere in the map, except within MMRs. This confirms that, in the case where H' is indeed constant, the vZLK mechanism is causing the rise in inclination we observed in Fig.~\ref{fig:NTA_aei}. It also shows that something else might be happening within the MMRs: they modify the constant of motion characteristic of the vZLK, as we will see in the next section. The MMRs we could identify are listed in Table~\ref{tab:MMR_all}. In order to investigate how the effects of the vZLK mechanism differ outside and within MMRs for Taurid particles, we also compared the evolution of our simulated particles with semi-analytical integral models of the long-term dynamics, both in and out of mean-motion resonances.

\begin{table}
    \caption{MMRs identified in the NTA BIN10100 map representing the variation of $H'$ }
    \label{tab:MMR_all}
    \centering
    \begin{tabular}{c c}
    \hline \hline
         MMR & $a$ (au) \\
         \hline
         6:1 & 1.58 \\
         5:1 & 1.78 \\
         9:2 & 1.91 \\
         4:1 & 2.06 \\
         11:3 & 2.19 \\
         7:2 & 2.26 \\
         10:3 & 2.33 \\
         3:1 & 2.50 \\
         8:3 & 2.70 \\
         \hline
    \end{tabular}
\end{table}

\subsubsection{Semi-analytical analysis of non-resonant vZLK dynamics}\label{subsubsec:nonres}

For particles outside mean-motion resonance, and assuming there is no short-timescale chaos, we can average the dynamics over the mean longitudes of both the particle and the perturbing planets. This averaging procedure amounts to a change of coordinates towards the secular (i.e. averaged) coordinates. In the secular coordinates, the semi-major axis $a$ of the particle is constant. Moreover, if we consider a somewhat idealized model of the Solar System in which all planets evolve on circular and coplanar orbits, the third component of the angular momentum of the particle is also constant, that is, the quantity $H'=\sqrt{1-e^2}\cos I$ is conserved over time, where $e$ and $I$ are the secular eccentricity and secular inclination of the particle (see \citealp{Saillenfest-etal_2016}).

Through the constant value of $H'$, any given value of $e$ corresponds to a given value of $I$. This link between $e$ and $I$ imposes a maximum value ever reachable by the particle, namely $e_\mathrm{max} = \sqrt{1 - H'^2}$, which is reached when $I=0$. The eccentricity and inclination ranges for the NTA and STA translate into values of roughly $H'\in[0.5,0.64]$.

\begin{figure*}
	\centering
	\includegraphics[width=0.4\textwidth]{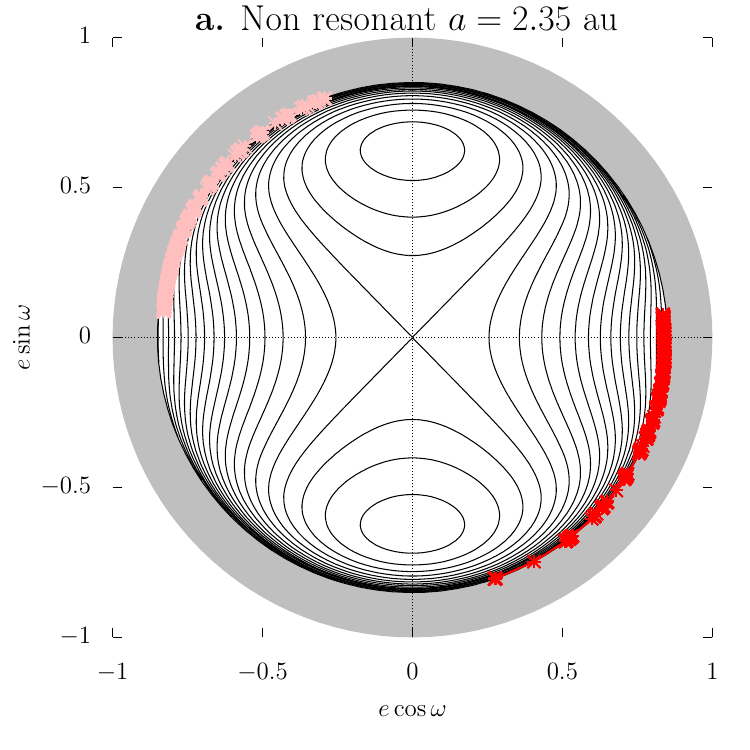}
	\includegraphics[width=0.4\textwidth]{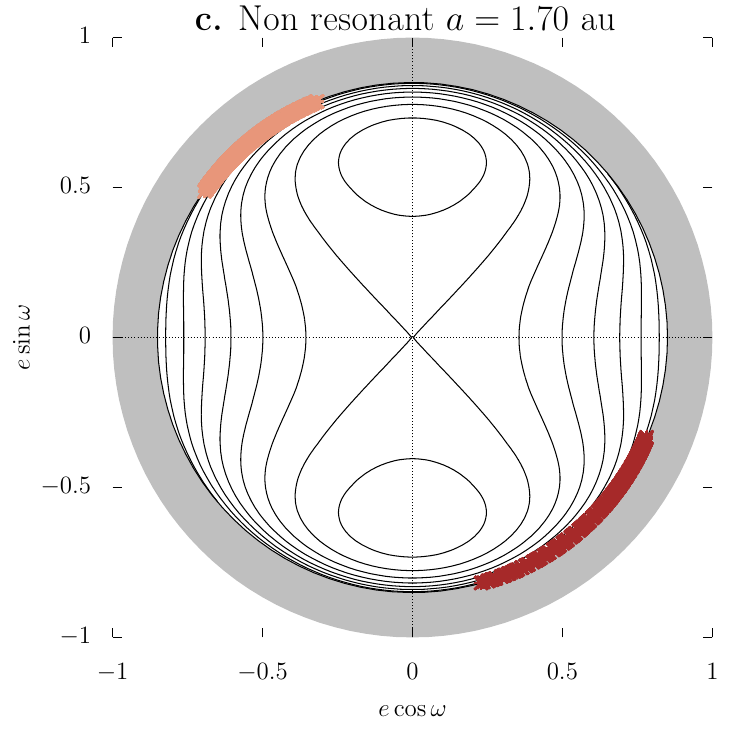}
	\\
	\includegraphics[width=0.4\textwidth]{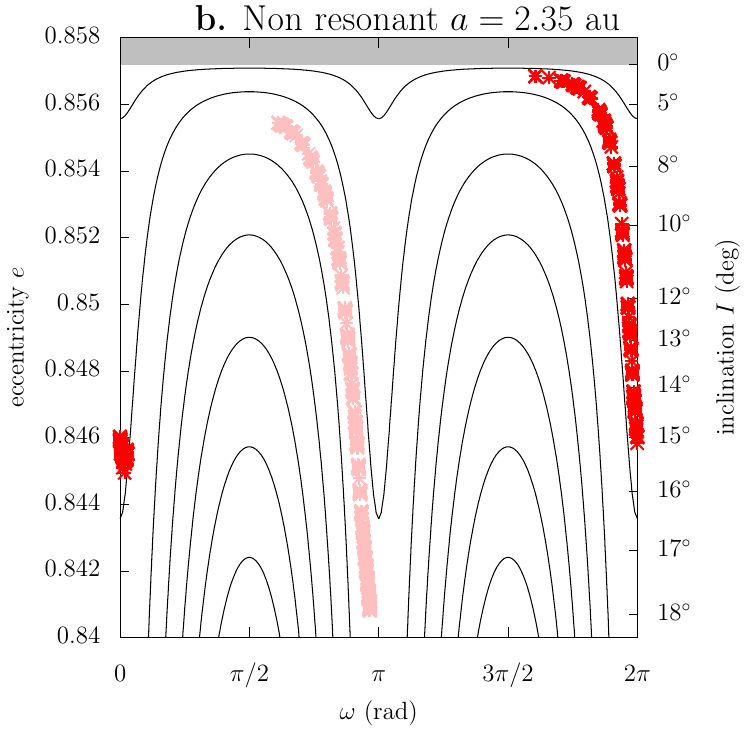}
    \includegraphics[width=0.4\textwidth]{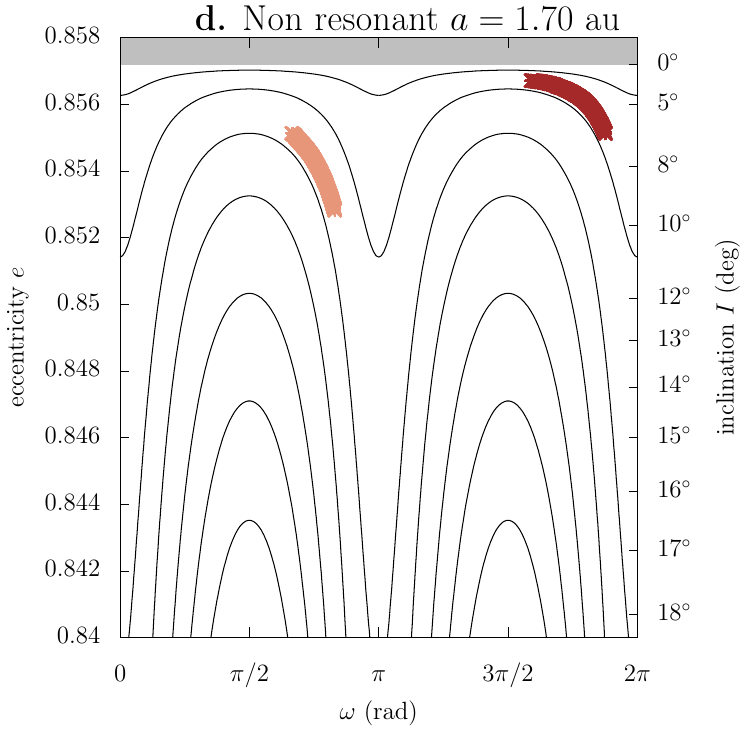}
	\caption{Integrable models for the non-resonant long-term dynamics of Taurid particles. The first case (panels~a, b) has parameters $a=2.35$~au and $H'=0.515$. The second case (panels~c, d) has parameters $a=1.70$~au. The gray region is forbidden for this value of $H'$. Top panels show a global polar view in the $(e\cos\omega,e\sin\omega)$ plane; bottom panels show enlarged views in the $(\omega,e)$ plane, with $I$ on the right axis linked to $e$ through the value of $H'$. Trajectories of particles from the NTA BIN10100 dataset (darker) and STA BIN10100 (lighter) are overplotted with colored points.}
	\label{fig:phaseportrait_nonres}
\end{figure*}

As we are left with only one degree of freedom, every possible trajectory for the particle can be represented by a level curve of the Hamiltonian function in the $(\omega,e)$ plane, with $a$ and $H'$ as parameters \citep{Thomas_Morbidelli_1996}. Panels~a and b of Fig.~\ref{fig:phaseportrait_nonres} show typical phase portraits obtained for non-resonant Taurids, computed here for $a=2.35$~au and $H'=0.515$. We also show an example of trajectory having similar $a$ and $H'$ from particles selected in our NTA BIN10100 dataset (coloured points).

The orbital inclination $I$ of particles in the N-body simulations follows a much smoother evolution than their eccentricity; therefore $I$ is more directly representative of their secular (i.e. averaged) dynamics than $e$. For this reason, we actually plot the coloured points in Fig.~\ref{fig:phaseportrait_nonres} in the $(\omega,I)$ plane (see right axis in panel~b), which is related to the secular value of $e$ through the $H'$ constant\footnote{Alternatively, one could also use filtering techniques to remove short-period oscillations from the output of N-body simulations and reveal their secular behaviour (see e.g. \citealp{Saillenfest-Lari_2017}). However, such a refined approach is not needed for the qualitative exploration conducted here.}.

The good agreement between phase portraits and N-body simulations shows that the secular model captures the essence of the long-term dynamics of simulated particles. The small deviations observed come from orbits of the planets in the N-body simulations which are not exactly circular and not exactly lying in the ecliptic plane, and from the orbit of the particle which is affected by slow chaotic diffusion.

Figure~\ref{fig:phaseportrait_nonres}a shows that Taurid particles are located outside the vZKL libration island; as such, their secular eccentricity variations are small. However, Fig.~\ref{fig:phaseportrait_nonres}b shows that these small eccentricity variations (left axis) translate into large oscillations in inclination (right axis), reaching more than $15^\circ$. For a smaller value of the semi-major axis $a$ used as parameter (Fig.~\ref{fig:phaseportrait_nonres}c, d), the oscillation cycles of Taurid particles have a smaller amplitude and they take longer to be completed, such that the inclination variations over 1000 years of Taurid particles having $a \approx 1.7$~au are much more modest than for those having $a \approx 2.35$~au (compare panels a, b with panels c, d).

\begin{figure}
    \includegraphics[scale = 0.6]{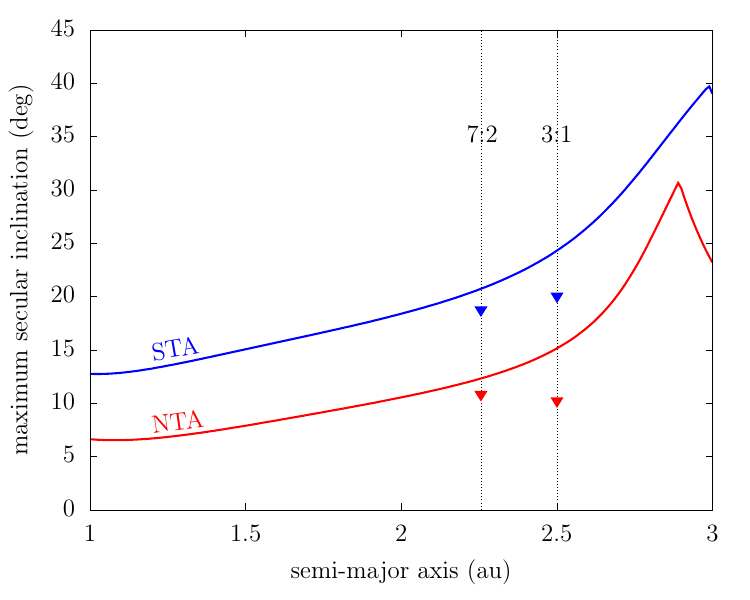}
    \caption{Maximum secular inclination reachable for particles depending on their initial semi-major axis. The triangles represent the maximum value of inclination when the particles are trapped in a mean-motion resonance.}
    \label{fig:evol_i}
\end{figure}

Fig.~\ref{fig:evol_i} shows the value of the secular inclination for $\omega = \pi$ on the level curve of the Hamiltonian, for the initial condition of the NTA and STA and various values of semi-major axis. This figure validates our previous interpretation: the maximum reachable inclination is much higher for larger semi-major axis. Furthermore, the maximum value of inclination reached by vZLK cycles within MMR is smaller than within MMR, as we will see in Sect.~\ref{subsubsec:res}. This figure ties together the chaos map (Fig.~\ref{fig:NTA_reso}), the maximum inclination map (Fig.~\ref{fig:NTA_aei}) and the variation of H' map (Fig.~\ref{fig:NTA_aeH}): the possible variation of inclination according to vZLK mechanism gets smaller as the semi-major axis gets smaller, which allows for the chaos to grow.

\subsubsection{Semi-analytical analysis of resonant vZLK dynamics}\label{subsubsec:res}

If the particle is trapped in a mean-motion resonance with a given planet, we can no longer average independently over the mean longitudes of both the particle and the planet involved. By averaging over fast angles, we obtain this time a semi-secular system which has two degrees of freedom \citep{Saillenfest-etal_2016}.

Thanks to the hypothesis that all planets have circular and coplanar orbits, there is still one conserved quantity, which can be written as $\gamma = \sqrt{a}(H' - k_p/k)$, where $k_p$ and $k$ are the coefficients of the resonance $k_p$:$k$ considered; for instance $k_p=3$ and $k=1$ for the $3$:$1$ resonance with Jupiter. As the particle is trapped in resonance by hypothesis, its semi-major axis $a$ oscillates about a central value $a_0$ which is characteristic of the resonance. For this reason, it is more convenient to replace the constant $\gamma$ by the equivalent constant $H'_0 = \gamma/\sqrt{a_0} + k_p/k$. Each time the particle's semi-major axis $a$ crosses $a_0$ during its oscillations, $H'_0$ is equal to $H'$. In other words, the constant $H'_0$ is equal to the mean value of $H'$. Therefore the parametrization by $H'_0$ (instead of $\gamma$) eases comparison with the non-resonant case (see paragraph~\ref{subsubsec:nonres}).

However, the system still has two degrees of freedom. The next step, in order to obtain an integrable model for the long-term dynamics inside the resonance, is to assume that the two degrees of freedom evolve on two very different timescales. This approximation is called the adiabatic approximation; it is more accurate for smaller planetary perturbations (\citealp{Lenard_1959,Henrard_1982,Wisdom_1985}; see also \citealp{Courtot_al_2024}). By virtue of this approximation, a new constant of motion emerges, $J$, which is equal to the area enclosed by the trajectory in the resonant domain of the phase space (see \citealp{Saillenfest_2020} for details). For instance, $J=0$ for zero-amplitude oscillations at the resonance center.

The constant $J$ removes one degree of freedom from the system, which leaves us with an integrable model for the long-term dynamics inside the resonance. Similarly to the non-resonance case, any possible trajectory can now be represented by a level curve of the Hamiltonian function in the $(\omega,e)$ plane, with $J$ and $H'_0$ as parameters. Moreover, $e$ is again linked to $I$ via the constant $H'_0$, in a very similar way as in the non-resonant case.

\begin{figure*}
	\centering
	\includegraphics[width=0.4\textwidth]{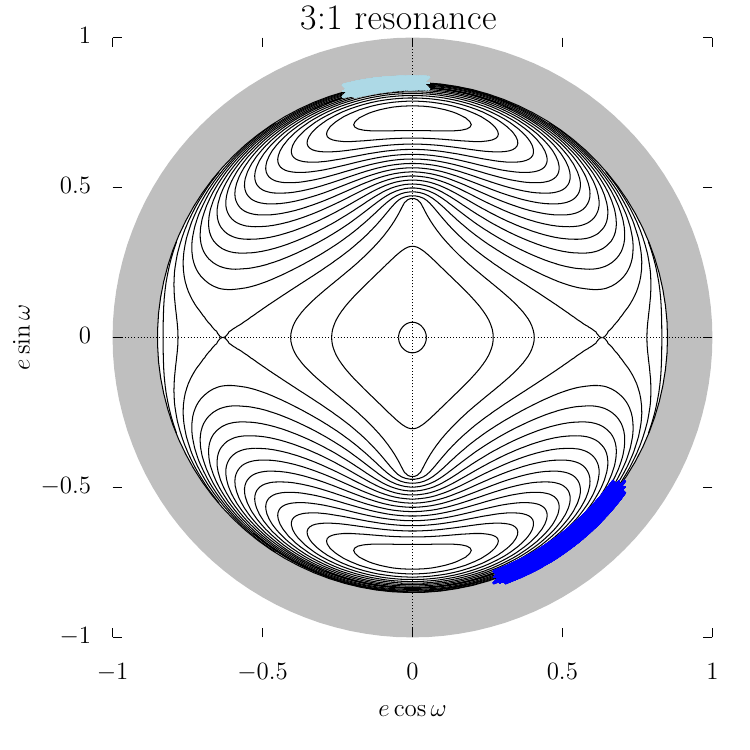}
	\\
    \includegraphics[width=0.4\textwidth]{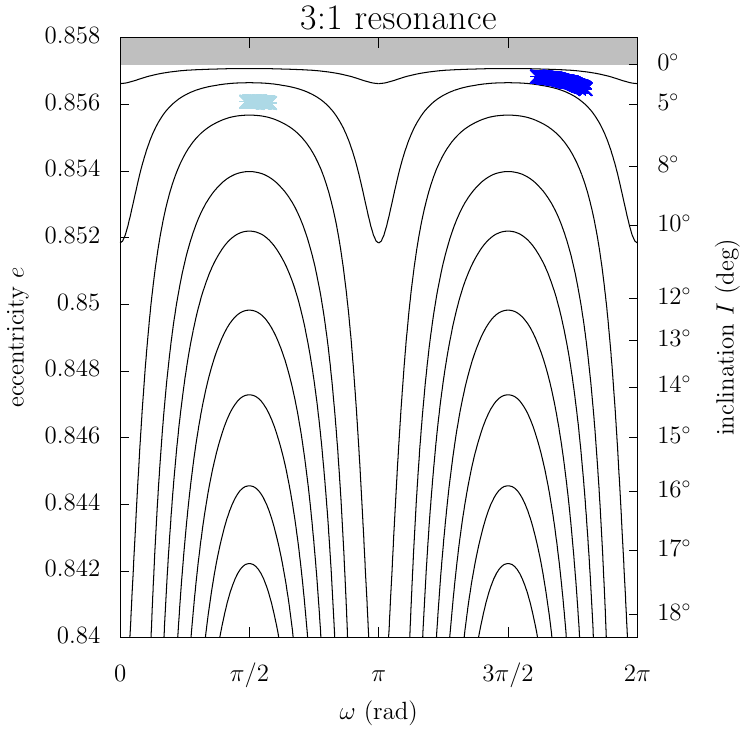}
	\caption{Integrable models for resonant long-term dynamics of Taurid particles within resonance 1:3 with $J=0$ and $H'_0=0.515$. The gray region is forbidden for this value of $H'_0$. Top panels show a global polar view in the $(e\cos\omega,e\sin\omega)$ plane; bottom panels show enlarged views in the $(\omega,e)$ plane, with $I$ on the right axis linked to $e$ through the value of $H'_0$. Trajectories of particles from the NTA BIN10100 dataset (darker) and STA BIN10100 (lighter) are overplotted with colored points.}
	\label{fig:phaseportrait_res}
\end{figure*}

Fig.~\ref{fig:phaseportrait_res} show typical phase portraits obtained for Taurids in the resonance $3$:$1$. As before, all Solar System planets are included and assumed to have circular and coplanar orbits; but now, a resonance is taken into account with Jupiter. The oscillation amplitude inside the resonance is assumed to be zero (i.e. $J=0$), which maximizes the influence of the resonance on the long-term dynamics (see \citealp{Saillenfest-etal_2017}). By comparing Fig.~\ref{fig:phaseportrait_nonres}a and the top panel of Fig.~\ref{fig:phaseportrait_res}, we notice the strong influence of the mean-motion resonance on the long-term dynamics of the particles: new equilibrium points appear, together with new vZLK libration islands.

Comparing Fig.~\ref{fig:phaseportrait_nonres}b and the bottom panel of Fig.~\ref{fig:phaseportrait_res}, we see that for similar parameters $H'$ and $H'_0$ and similar initial conditions, resonant particles undergo smaller variations in eccentricity and inclination than non-resonant particles, just as a result of the geometry of the phase space. Those lower maximum values of inclination are represented in Fig.~\ref{fig:evol_i} by the small triangles. Moreover, oscillation cycles of resonant particles in this region of the phase space are slower compared to non-resonant particles. These two factors (the geometry and the timescale) contribute to produce less variations in inclination over 1000~years for resonant particles than for non-resonant particles.

\subsection{A deeper analysis of 7:2 Jupiter}
In the past, the 7:2 MMR with Jupiter has been cited as responsible for some specificity in the Taurids observations. In particular, it could lead to a clustering of particles that would be detected as a swarm \citep{Asher_Izumi_1998}. Thus, we have also analysed this MMR in further details. It is clearly present in our maps, as we have shown before, but it is much thinner than the other ones, due to its higher order.

Particles inside the 7:2 MMR with Jupiter have a resonant angle $\sigma$ that librates around 0\degr. This angle is defined as: $\sigma = 7\lambda_J - 2\lambda - 5\varpi$, with $\lambda$ the mean longitude of the resonant particles, $\lambda_J$ the mean longitude of Jupiter and $\varpi$ the perihelion longitude of the resonant particles. We obtain the equation: $2M = 7\lambda_J-7\varpi$, with $M$ the mean anomaly of the resonant particles. If $\varpi$ is almost constant for all the particles, then there is two solutions of this equation for $M$, separated by 180\degr. This will be visible as two peaks in the mean anomaly distribution. \citet{Asher_Izumi_1998} show that one of these peaks is visible in the observations, explaining that the parent body librates around one of the possible values for $M$ and thus the observed Taurids do the same.

\begin{figure}
    \includegraphics[scale = 0.5]{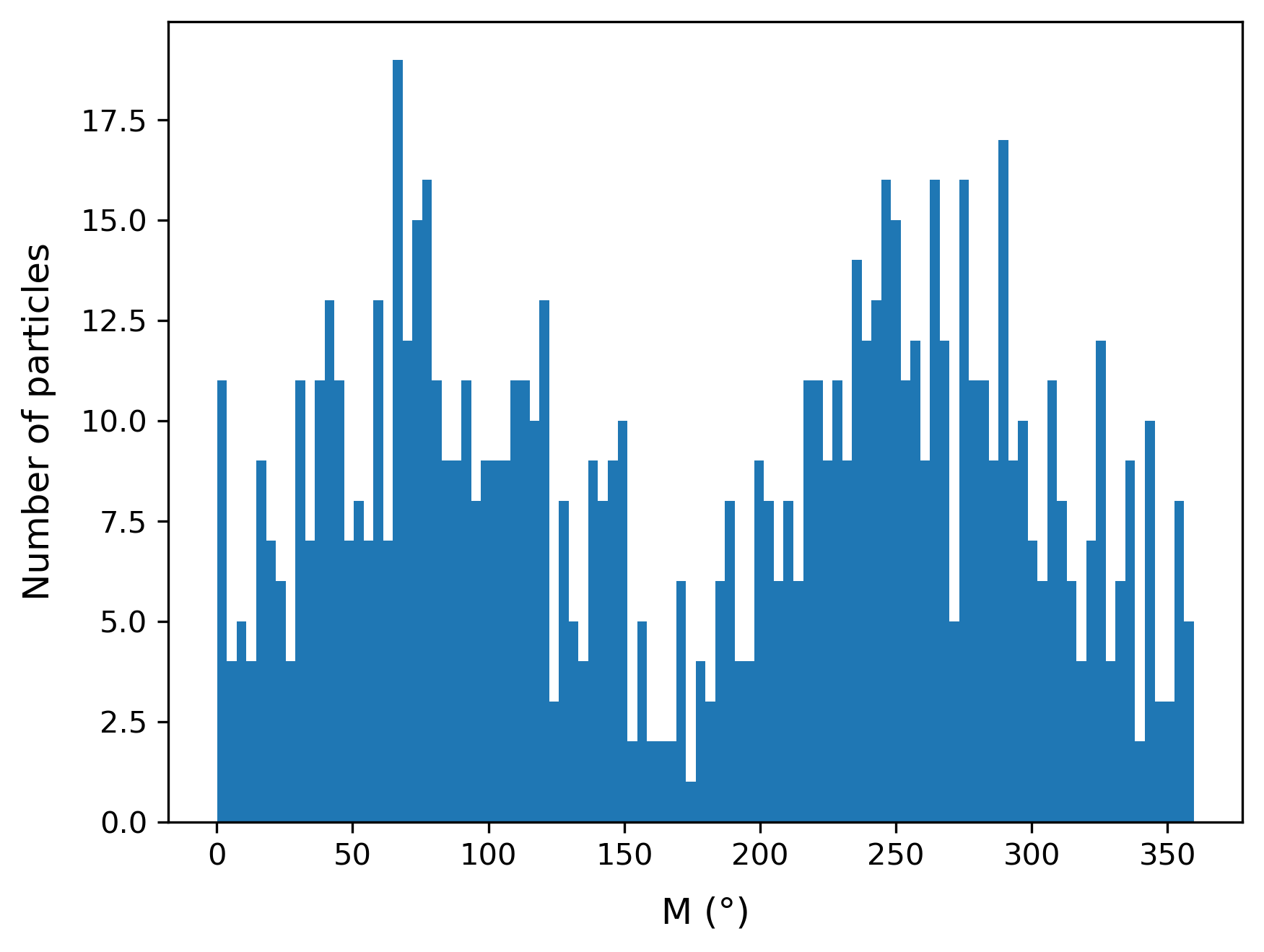}
    \caption{Distribution (100~bins) of the final mean anomaly of particles inside the 7:2 MMR with Jupiter and with a $\varpi$ close to 165\degr (see text for precise description of the selection process).}
    \label{fig:hist_M}
\end{figure}

To check for this result in our simulation, we selected particles that are inside the MMR by making sure their inclination stays inferior to ten~degrees through the integration and their initial semi-major axis $a$ follows $|a - 2.26| < 0.05$. Then we plotted the distribution in final longitude of periapsis $\varpi$ of these particles and found there was enough particles with $|\varpi - 165|$\degr$ < 10$\degr to plot a significant histogram. Fig.~\ref{fig:hist_M} shows the distribution of final mean anomaly for these particles:~we do see two peaks that correspond to \citet{Asher_Izumi_1998} results.

\begin{figure*}
	\centering
	\includegraphics[width=0.4\textwidth]{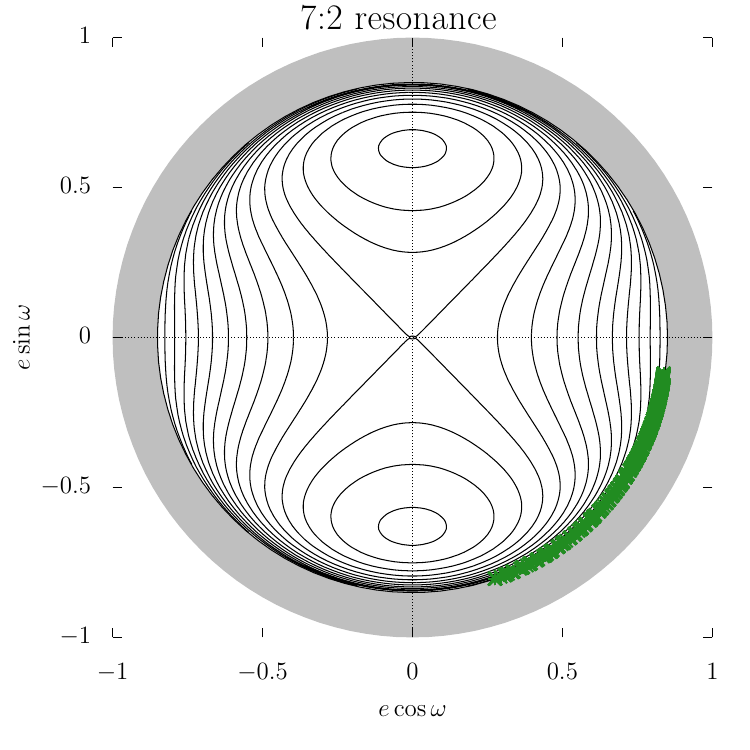}
    \\
    \includegraphics[width=0.4\textwidth]{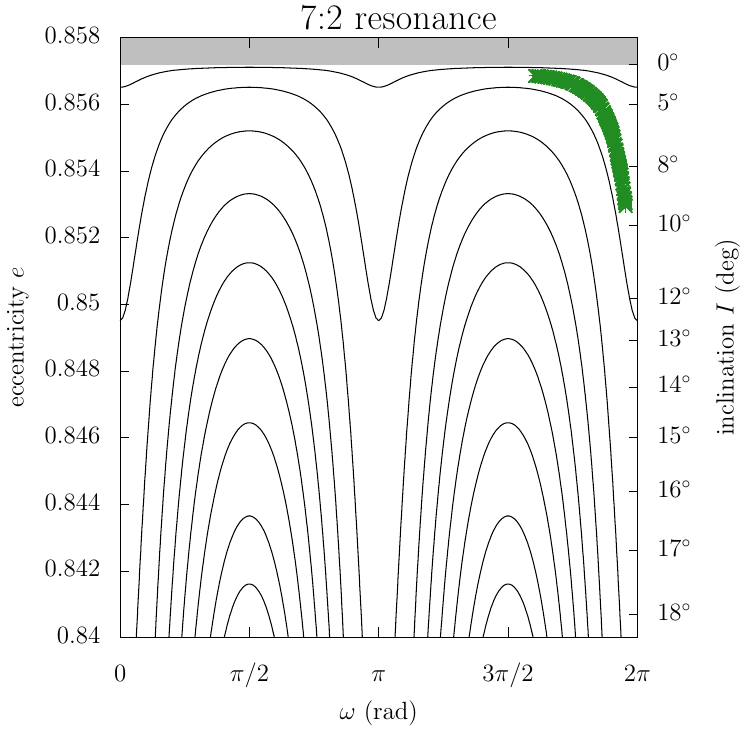}
	\caption{Integrable models for the resonant long-term dynamics of Taurid particles. The model have parameters $J=0$ and $H'_0=0.515$. The gray region is forbidden for this value of $H'_0$. The top panel shows a global polar view in the $(e\cos\omega,e\sin\omega)$ plane; the bottom panel show enlarged views in the $(\omega,e)$ plane, with $I$ on the right axis linked to $e$ through the value of $H'_0$. Trajectories of particles from the NTA BIN10100 dataset are overplotted with coloured points.}
	\label{fig:phaseportrait_72}
\end{figure*}

Finally, we also analyse the vZLK mechanism regarding the 7:2 MMR with Jupiter. Following the method described in paragraph~\ref{subsubsec:res}, we present the model for resonant particles in 7:2 with Jupiter, with some trajectories of particles selected from the NTA BIN10100 overplotted (see Fig.~\ref{fig:phaseportrait_72}). Comparing this figure with Fig.~\ref{fig:phaseportrait_res}, we note that the $7$:$2$ resonance distorts much less the phase space than the stronger $3$:$1$ resonance. As a result, long-term dynamics within the $7$:$2$ resonance qualitatively resembles non-resonant dynamics (see Fig.~\ref{fig:phaseportrait_nonres}a and Fig.~\ref{fig:phaseportrait_72}a); yet, inclination variations are still less pronounced within the resonance than outside (compare Fig.~\ref{fig:phaseportrait_nonres}b and Fig.~\ref{fig:phaseportrait_72}b). This is represented as the small triangle in Fig.~\ref{fig:evol_i}. Again the geometry of this portion of phase space and the evolution timescale contribute to produce less orbital variations over 1000~years for particles in the  $7$:$2$ resonance than for non-resonant particles.

This explains why the 7:2 MMR with Jupiter is indeed detected in the maps, while being much thinner than other resonances.

\subsection{Non-gravitational forces}\label{subsec:fng}

Our dataset NTA BIN10100 minimizes the effect of non-gravitational forces owing to the large size of the particles. In this section, we focus on the NTA BIN00101 dataset, where the non-gravitational forces (Poynting-Robertson drag and solar radiation pressure) are maximised. By comparing these results, we explore how these forces modify the mechanisms we identified (MMRs and vZLK).

\begin{figure}
    \includegraphics[scale = 0.5]{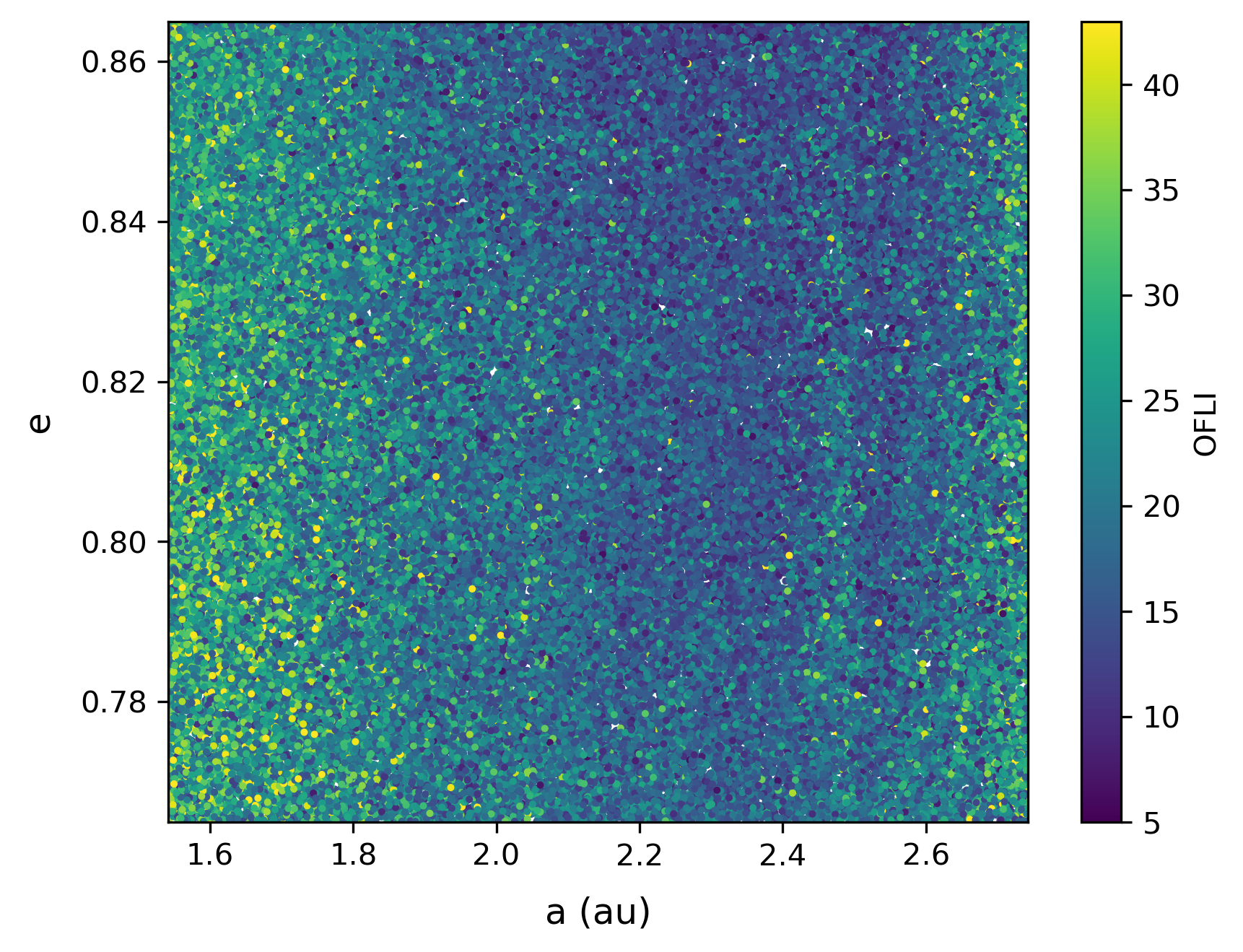}
    \caption{Semi-major axis and eccentricity chaos map of the NTA BIN00101. For easier comparison with previous map, the colorbar is once again restricted to values between 5 and 43.}
    \label{fig:NTA_small}
\end{figure}

Fig.~\ref{fig:NTA_small} shows the chaos map we obtained from integrating the NTA BIN00101 dataset. The colorbar is also limited between 5 and 43, in order to ease the comparison with previous maps. Only 0.5\% of particles reach a final OFLI above 43, with a maximum of 66.66. This is slightly higher than the NTA BIN10100, but not significantly. The MMRs are not visible anymore, except maybe for the 3:1 with Jupiter, visible as a faint line with a higher OFLI. The dark area in the middle, previously due to the vZLK mechanism, is still present.

The Poyting-Robertson drag is the non-gravitational force responsible for the diminution of the semi-major axis of the small particles \citep{Liou_Zook_1997}, which can make them escape from the MMRs. We verify this is what happening here by computing the drift in semi-major axis for our particles thanks to the equation:
\begin{equation}\label{eq:drift}
    \left( \frac{da}{dt} \right)_{PR} = \frac{-1.35 \beta \mu_\sun}{c} \frac{2 + 3 e^2}{a (1 - e^2)^{3/2}},
\end{equation}
with $\beta$ the ratio of the radiation pressure force and the Sun gravitational force, $\mu_\sun$ the gravitational constant of the Sun and $c$ the speed of light. 

We also have \citep{Burns_al_1979}:
\begin{equation}
    \beta = \frac{3}{4} \frac{S_0 R_0^2}{\mu_\sun c} \frac{1}{\rho r},
\end{equation}
with $\rho$ the bulk density of the particle, $r$ its radius and $S_0$ the solar flux at $R_0 = 1$~au (its value is $1.37 \text{kW}\ \text{m}^{-2}$ \citep{Cox_2000}). 

For a radius of 0.01~mm, the drift is equal to $-1.61 \times 10^{-11}$ au/s, which equivalent to -0.51 au over the whole integration. Thus such small particles escaped all the MMRs identified in the map.

\begin{figure}
    \includegraphics[scale = 0.5]{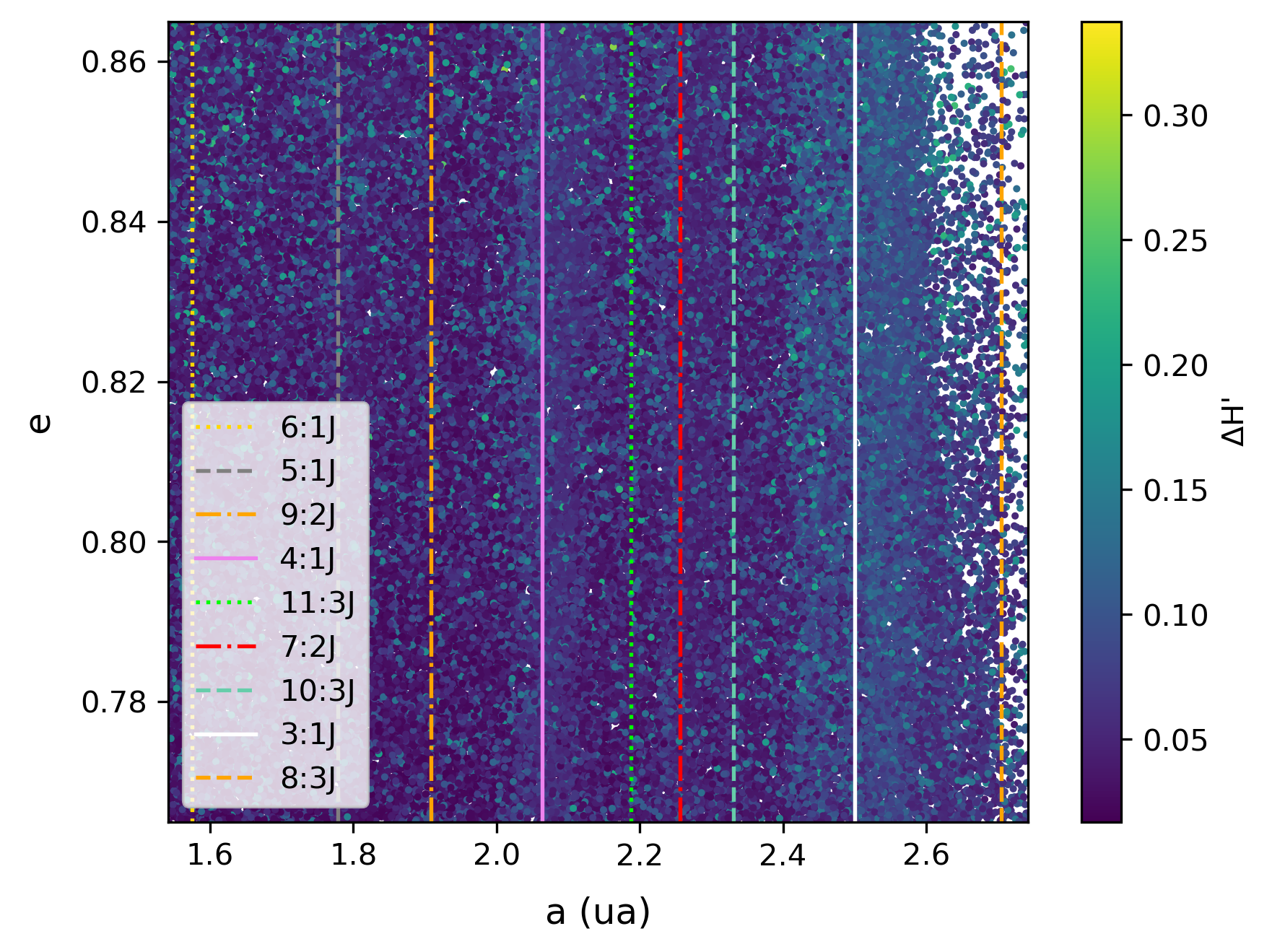}
    \includegraphics[scale = 0.5]{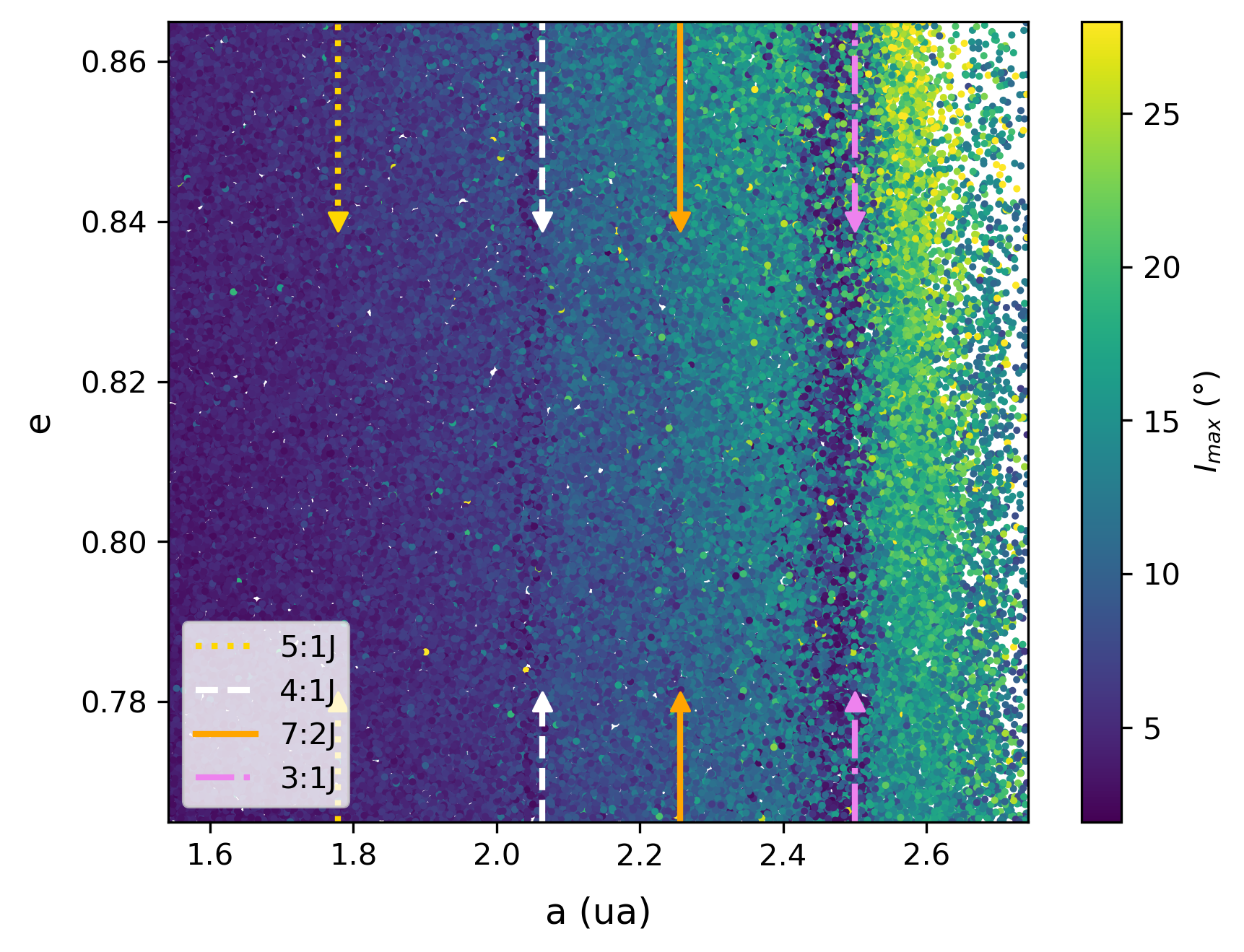}
    \caption{Variation of $H'$ (first figure) and maximum inclination reached during the integration (second figure) plotted in function of the initial semi-major axis and eccentricty. Only particles that do not encounter Jupiter are plotted.}
    \label{fig:NTA_small_Lidov}
\end{figure}

To verify if the vZLK mechanism is still responsible for the dark area, we plot again the variation in $H'$ and the maximum inclination reached during the integration in Fig.~\ref{fig:NTA_small_Lidov}. Two MMRs do play a role in the evolution of the NTA BIN00101 particles: the 3:1 with Jupiter and, although almost negligible, the 4:1 with Jupiter. This makes sense, as these MMRs are the widest resonances in the dynamics. It seems these MMR are strong enough to prevent the inclination from raising too much during the integraion, but not enough for it to make a clear difference in the chaos map. This map also proves the vZLK mechanism plays the same role for the small particles despite the drift in semi-major axis. This drift means the semi-analytical model is not valid anymore, and makes such analysis impossible, as this dynamic is non-hamiltonian. However, the figure presented here is clear enough to conclude on the validity of the vZLK mechanism identified previously.

We also computed the drift for a radius of 0.3~mm, the value we chose for our computation of the Geminids and Draconids drift. We obtain a value of -0.017~au over the whole integration, which is smaller than the Geminids (-0.048~au over 1000~years) and larger than the Draconids (-0.006~au over 1000~years). The difference between the Geminids drift and the NTA drift is especially telling, since their semi-major axis is more similar to the NTA than the Draconids. The Poynting-Robertson drag is much stronger for the Geminids and the effect of the vZLK is also negligible for this stream, setting the NTA apart.

\subsection{Southern vs Northern Taurids}\label{subsec:STAvsNTA}

\begin{figure}
    \includegraphics[scale = 0.5]{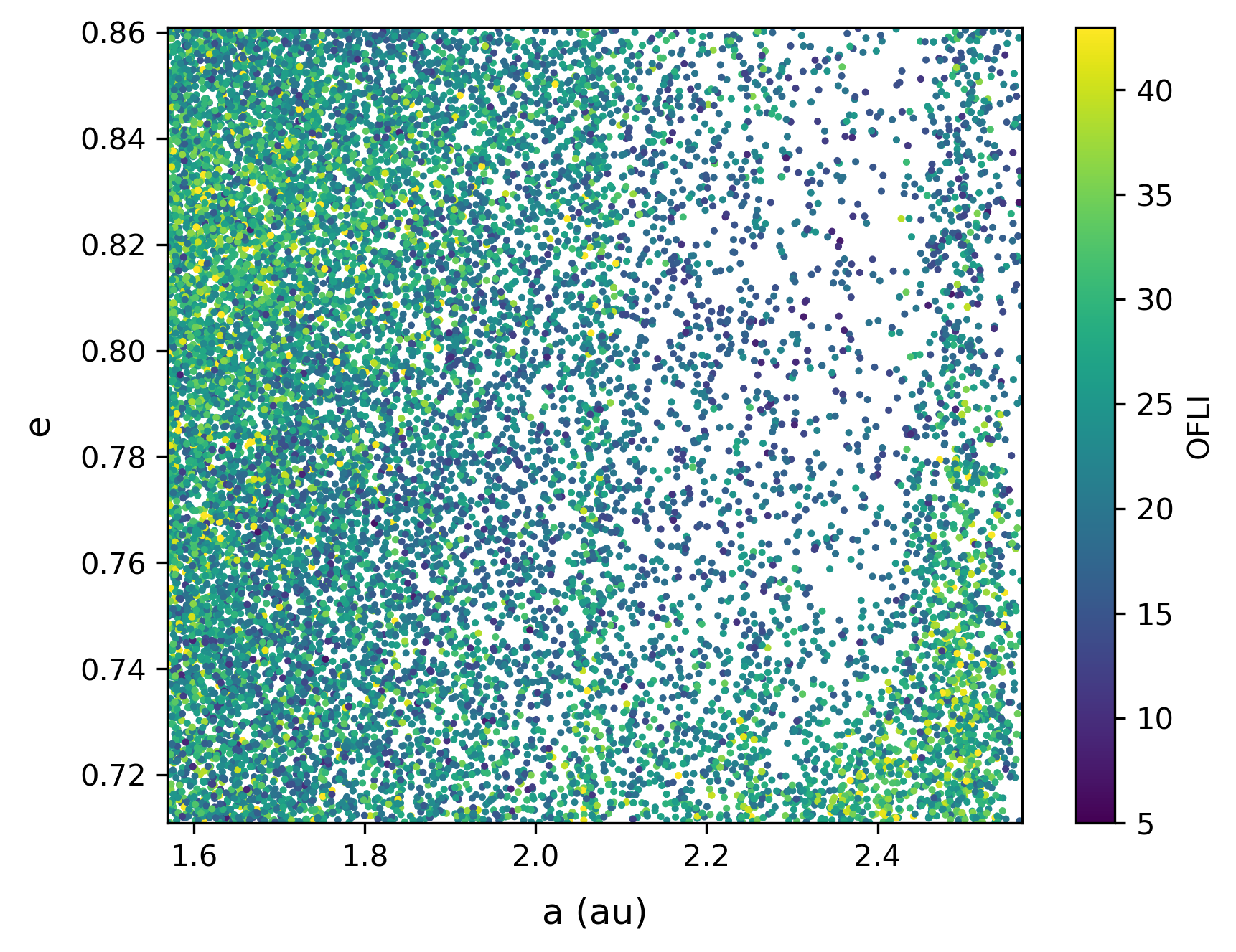}
    \caption{Semi-major axis and eccentricity chaos map for the STA BIN10100. Only particles that encounter the Earth, Mars or Venus are plotted.}
    \label{fig:STA_renc}
\end{figure}

Finally, we analysed the STA to see if similar mechanisms can be identified. First, Fig.~\ref{fig:STA_renc} shows the particles that encounter the Earth, Mars or Venus during the integration. Both the amount of encounters and their distribution over the map seems similar. The only difference is the amount of encounters in the top left and bottom right corners, which is higher and explains the higher amount of chaos compared to the NTA. They are also linked with higher level of chaos for the smaller initial semi-major axis. Contrary to the NTA, only 42~particles encounter Jupiter, as the range of the initial semi-major axis of the STA does not reach sufficiently high value, contrary to the NTA. 

\begin{figure}
    \includegraphics[scale = 0.5]{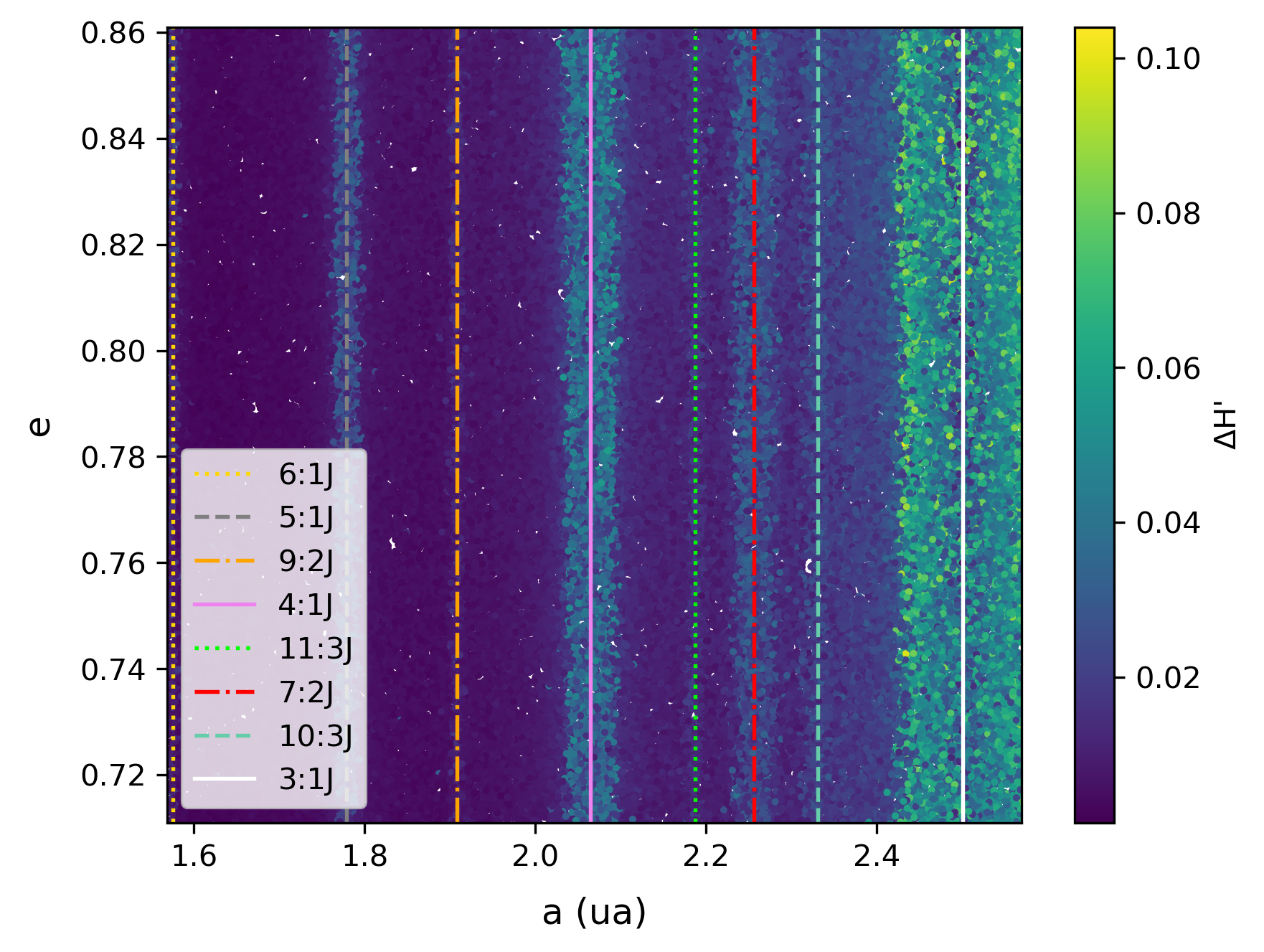}
    \includegraphics[scale = 0.5]{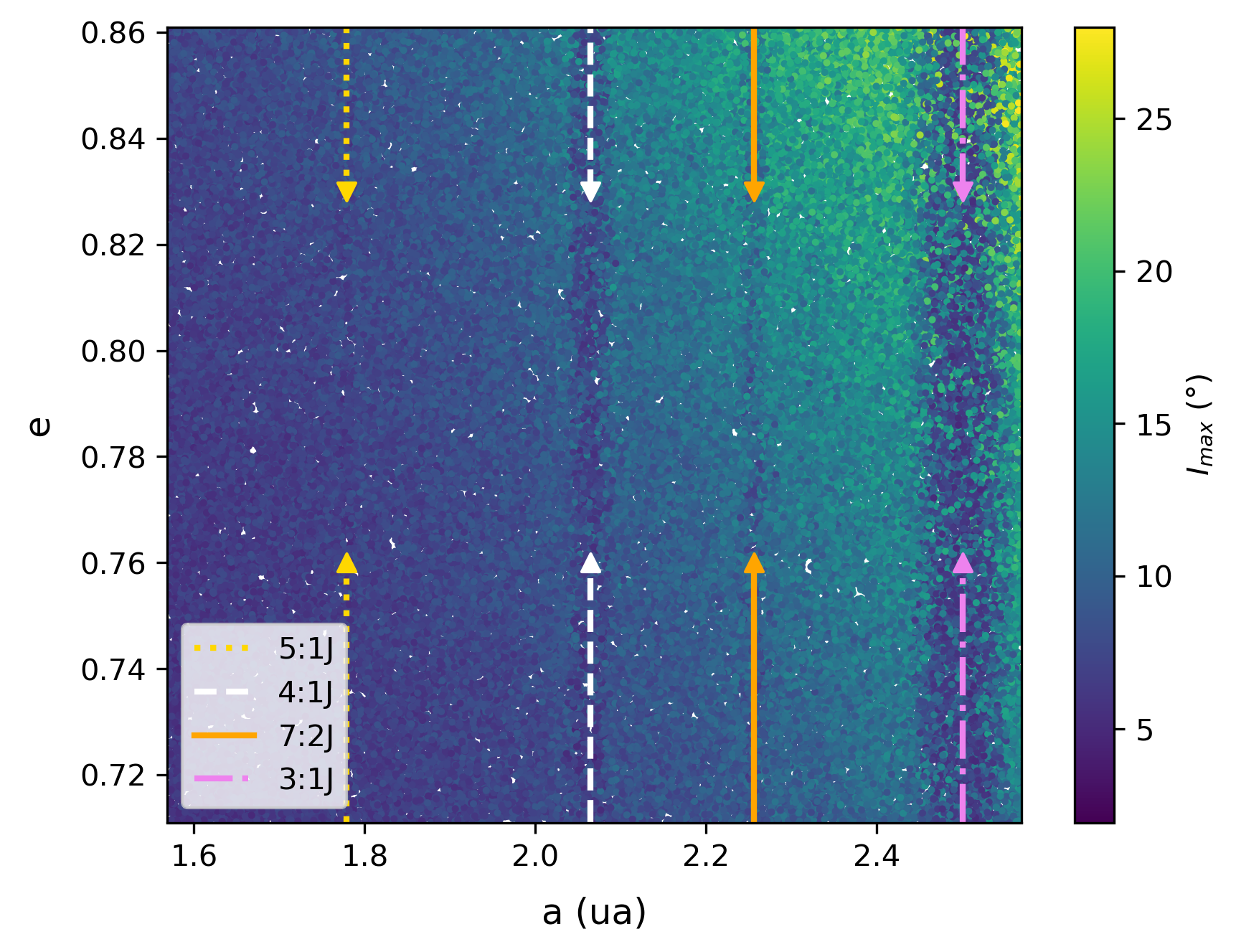}
    \caption{For the dataset STA BIN10100, variation of $H'$ (first figure) and maximum inclination reached during the integration (second figure) plotted in function of the initial semi-major axis and eccentricty.}
    \label{fig:STA_Kozai}
\end{figure}

We also verify the effect of the vZLK mechanism with Fig.~\ref{fig:STA_Kozai}. Once again, the same MMRs play the same role for the STA as the NTA - capturing the particles and preventing them for reaching a higher inclination while the rest of the particles reach higher inclination due to the vZLK mechanism.

The semi-analytical computations previously performed already included the STA. As we can see in Fig.~\ref{fig:phaseportrait_nonres} and Fig.~\ref{fig:phaseportrait_res}, the STA behave very similarly to the NTA in regards to the vZLK cycle. In the specific case of $a = 2.35$~au (Fig.~\ref{fig:phaseportrait_nonres}a and b), we note that the oscillation cycle for the STA particles is too long to actually reach its maximum over the 1000~year duration of the integration. 

This validates our initial hypothesis that the dynamics between NTA and STA are substantially identical. Apart from the lack of close encounters with Jupiter, the STA behave in the same way as the NTA, as their orbits are not different enough to experience different dynamics.

\section{Conclusion}\label{sec:concl}

Our chaos maps allow us to describe in details the dynamics of the Taurids. The semi-analytical analysis confirm these results.

Close encounters with Jupiter, for particles in the NTA with a high initial semi-major axis, disrupt the dynamics significantly, while close encounters with the Earth, Venus and Mars, for particles with a small initial semi-major axis, can also bring more chaos to the particles.

The vZLK mechanism links the eccentricity and inclination. The specific geometry of the orbits of the NTA and STA means a small variation in eccentricity translates to a high variation in inclination. This means that most particles get further from the ecliptic plane, which changes the amount of encounters and the encounters geometry, maintaining them in a regular dynamic.

This is especially true for the particles with a larger semi-major axis within the Taurid streams, as this change in inclination gets smaller for particles with a semi-major axis smaller than about 1.7~au.

MMRs change this picture: the long-term dynamics within the MMRs we identified result in less inclination variations and these variations are also slower to develop. This means that particles trapped in the MMRs tend to remain longer near the ecliptic plane and thus their chaoticity level rises. This is a very different picture from what we observed for the Geminids, the Draconids and the Leonids, where MMRs preseved the regularity of the orbits.

Non-gravitational forces make this picture more complex for the smallest particles: the drift in semi-major axis due to the Poynting-Robertson drag allows for the smallest particles to escape the MMRs and be swept up in the vZLK mechanism.

The picture our chaos maps paint is completely different from what we found for the Geminids, Draconids and Leonids. First, none of these stream experienced the vZLK mechanism to the extent the NTA and STA do. Second, the value of the drift in semi-major axis created by the Poyting-Robertson drag sits between the Geminids and the Draconids, allowing some MMRs to retain a small effect. It is tempting to explain the difficulty in finding the origin of the NTA and STA by the particularity of these two streams compared to the other three, but further analysis and comparisons are necessary for such a claim to be made. However, we can already say that most observations of particles with a semi-major axis lower than 1.8~au or higher than 2.4~au might be difficult to trace back to a parent body, owing to the high level of chaos experienced by most particles in this area.

In the future, we would also like to confront our maps with the observational data. Indeed, we were able to find in our simulations a confirmation of the observed mean anomaly peak for particles inside the 7:2 MMR with Jupiter, which could be responsible for the Taurids swarms, but other observational characteristics were not studied here.

\begin{acknowledgements}
We thank David Asher for some nice discussion on the Taurids and the 7:2-related swarms.
\end{acknowledgements}

\bibliographystyle{aa} 
\bibliography{biblio.bib}

\end{document}